\documentclass[11pt]{article}

\usepackage{amsmath, amssymb, amsthm} 
\usepackage{graphicx} 
\usepackage{hyperref} 
\usepackage{geometry} 
\usepackage{cite} 
\usepackage{setspace} 

\usepackage{subcaption}
\usepackage{xcolor}  
\usepackage{tcolorbox}

\title{A Hybrid Framework for Reinsurance Optimization: Integrating
Generative Models and Reinforcement Learning}

\author{
Stella C. Dong \\
Reinsurance Analytics \\
\texttt{stella.dong@reinsuranceanalytics.io}
}

\date{}

\usepackage{tikz} 

\usetikzlibrary{arrows.meta, positioning, decorations.pathmorphing}
\usetikzlibrary{shapes}
\usetikzlibrary{calc} \tikzset{>=latex}
\usetikzlibrary{positioning,arrows, arrows.meta,bending,automata, patterns,snakes, decorations.pathreplacing, shadows, calligraphy}
\usetikzlibrary{arrows.meta, positioning, shapes.geometric}

\usepackage{amsthm}

\usepackage{tabularx} 
\numberwithin{equation}{section}

\usepackage{listings}
\usepackage{algpseudocode}
\usepackage{algorithm}
\usepackage{tcolorbox}
\usepackage{hyperref}

\begin{document}

\maketitle

\begin{abstract}
Reinsurance optimization is a cornerstone of solvency and capital management, yet traditional approaches often rely on restrictive distributional assumptions and static program designs. 
\textcolor{black}{We propose a hybrid framework that combines Variational Autoencoders (VAEs) to learn joint distributions of multi-line and multi-year claims data with Proximal Policy Optimization (PPO) reinforcement learning to adapt treaty parameters dynamically.} 
\textcolor{black}{The framework explicitly targets expected surplus under capital and ruin-probability constraints, bridging statistical modeling with sequential decision-making.} 

\textcolor{black}{Using simulated and stress-test scenarios---including pandemic- and catastrophe-type shocks---we show that the hybrid method produces more resilient outcomes than classical proportional and stop-loss benchmarks, delivering higher surpluses and lower tail risk.}  
\textcolor{black}{Our findings highlight the usefulness of generative models for capturing cross-line dependencies and demonstrate the feasibility of RL-based dynamic structuring in practical reinsurance settings.}  

\textcolor{black}{Contributions include (i) clarifying optimization goals in reinsurance RL, (ii) defending generative modeling relative to parametric fits, and (iii) benchmarking against established methods. This work illustrates how hybrid AI techniques can address modern challenges of portfolio diversification, catastrophe risk, and adaptive capital allocation.}
\end{abstract}


\section{Introduction}
\label{sec:introduction}

The insurance and reinsurance industries play a pivotal role in managing financial risks and ensuring economic stability. Reinsurance, which involves the transfer of risk from insurers to reinsurers, is a cornerstone of risk management strategies aimed at maintaining solvency and optimizing financial performance. However, designing effective reinsurance strategies remains a highly complex challenge due to the stochastic nature of claims, multi-dimensional constraints, and the dynamic interplay between risk retention, profitability, and regulatory compliance \cite{Asmussen_and_Albrecher2010, Schmidli2008}. 

\textcolor{black}{A central difficulty lies in managing tail risk: catastrophic but rare claims can dominate portfolio outcomes, and classical actuarial models often underestimate such events. This motivates the need for adaptive, data-driven frameworks that explicitly account for tail behavior and align with evolving solvency regulations.}\textcolor{black}{This emphasis on catastrophe risk and model-based stress testing aligns with recent RMIR perspectives on data science for catastrophe modeling and disaster risk reduction \cite{Steptoe2022_RMIR,Surminski2018_RMIR}.}

Traditional approaches to reinsurance optimization, such as the classical Cram\'{e}r---Lundberg model, have provided foundational insights into surplus dynamics and ruin probabilities. These models, while mathematically rigorous, rely on static assumptions about premium rates and claim distributions, limiting their applicability to modern reinsurance practices. Extensions to these models, including proportional and layered reinsurance structures, address some of these limitations but often remain computationally intensive and insufficiently adaptable to high-dimensional, real-world scenarios \cite{Gerber1970, Albrecher_et_al2017}. \textcolor{black}{Recent surveys highlight that static optimization struggles in the presence of heavy-tailed claims and under regulatory frameworks such as Solvency~II and IFRS~17 \cite{Mikosch2009, Embrechts_et_al2014}.}

\textcolor{black}{Building on this, a growing literature explores optimization of reinsurance treaties in dynamic and stochastic settings \cite{Schmidli2008, Albrecher_et_al2017}. Parallel advances in machine learning open new opportunities: generative models such as Variational Autoencoders (VAEs) capture complex data distributions and generate synthetic samples, helping address data scarcity and the underrepresentation of catastrophic claims \cite{Kingma_and_Welling2014, Goodfellow_et_al2014, Wuthrich2020}. Reinforcement learning (RL), particularly Proximal Policy Optimization (PPO), has demonstrated strong performance in sequential decision-making under uncertainty \cite{Schulman_et_al2017, Kolm_et_al2020}, with recent applications highlighting its promise in insurance risk management \cite{Buehler2019, Buehler2022}.} \textcolor{black}{Related RMIR contributions underscore the role of advanced analytics and organizational learning for risk management, reinforcing the relevance of our hybrid approach \cite{TextMining2024_RMIR,Resilience2023_RMIR}.}

\textcolor{black}{Despite these advances, little work has combined generative modeling of claim processes with dynamic optimization of reinsurance strategies. Our contribution is to close this gap by proposing a hybrid framework that explicitly targets tail-risk robustness in reinsurance optimization.}

This paper introduces a novel hybrid framework that integrates generative AI with reinforcement learning to optimize reinsurance strategies dynamically and adaptively. By leveraging VAEs to model claim distributions and generate synthetic scenarios, the framework overcomes challenges associated with data scarcity and variability. The PPO algorithm, on the other hand, dynamically adjusts reinsurance parameters---such as retention rates and layer boundaries---based on evolving claim distributions, market conditions, and regulatory constraints. This synergy enables the framework to evaluate and optimize complex reinsurance strategies in real time, addressing high-dimensional uncertainties and ensuring financial stability \cite{Albrecher_et_al2017, Cheng_et_al2020}.

\textcolor{black}{The framework is empirically evaluated on three representative claim distributions---lognormal, Pareto, and a lognormal---Pareto mixture---selected for their relevance to insurance modeling and their contrasting tail behaviors \cite{Embrechts_et_al2014}. Results demonstrate improved robustness in the upper tail and closer alignment with solvency requirements compared to classical optimization approaches.}

Key contributions of this work include:
\begin{enumerate}
    \item \textbf{A hybrid framework for reinsurance optimization:} The integration of generative AI models (VAEs) and reinforcement learning (PPO) to address multi-dimensional, stochastic optimization challenges in reinsurance.
    \item \textbf{Dynamic parameterization of reinsurance strategies:} Incorporation of adaptive retention rates and layer boundaries to ensure flexibility in risk-sharing mechanisms under evolving market conditions.
    \item \textbf{Comprehensive validation across distributions:} \textcolor{black}{Empirical evaluation using lognormal, Pareto, and mixture distributions, demonstrating improved tail robustness and solvency alignment relative to established baselines.}
\end{enumerate}

{\textcolor{black}{The remainder of this paper is organized as follows. Section~\ref{sec:model_description} presents the mathematical foundations of the surplus process, reinsurance structures, and optimization objectives. Section~\ref{sec:hybrid_framework} introduces the proposed hybrid framework, describing the integration of generative modeling with reinforcement learning. Section~\ref{sec:experiments} details the experimental design, results, and benchmarking against established baselines. Section~\ref{sec:discussion} examines practical implications and limitations, while Section~\ref{sec:conclusion} summarizes key findings and outlines directions for future research. By uniting actuarial rigor with modern AI techniques, this study advances a new paradigm for reinsurance optimization---enhancing financial resilience, strengthening decision-making under uncertainty, and laying the groundwork for next-generation risk management strategies.}}

\section{Model Description}
\label{sec:model_description}

\textcolor{black}{This section presents the mathematical foundations of our framework, which is designed to capture the operations of an insurer over a finite planning horizon $T$}. \textcolor{black}{The framework combines a discrete-time formulation \cite{Asmussen_and_Albrecher2010}, a generalized surplus process \cite{Schmidli2008}, and flexible reinsurance mechanisms \cite{Gerber1970, Albrecher_et_al2017} to address the dual challenges of financial stability and risk management under uncertainty.} \textcolor{black}{By explicitly structuring the surplus process to accommodate both proportional and layered reinsurance treaties, as well as dynamic treaty adjustments, the model provides a tractable yet versatile basis for optimization \cite{Mikosch2009, Embrechts_et_al2014}.} 

\textcolor{black}{This mathematical foundation supports the integration of learning-based approaches in later sections. In particular, reinforcement learning agents optimize reinsurance decisions by adapting retention rates and layer structures to evolving claims and market conditions, extending classical actuarial models toward adaptive, data-driven frameworks \cite{Buehler2019, Wuthrich2020}.}

\subsection{Discrete-Time Framework}

The planning horizon $T$ is partitioned into $n$ discrete intervals, denoted $t_1, \ldots, t_n$, where $t_1 = 0$ and $t_n = T$. \textcolor{black}{Each interval serves as a decision epoch at which the insurer updates its risk portfolio, collects premiums, and settles claims. This setup mirrors industry practice, where financial positions are reviewed and adjusted at regular reporting periods, such as quarterly or annual solvency assessments \cite{Asmussen_and_Albrecher2010, Kaas2008}.}

\textcolor{black}{Formulating the model in discrete time enables fine-grained analysis of both risk exposure and financial stability. In particular, it facilitates the incorporation of stochastic variability in claims and premiums, while preserving tractability for optimization \cite{Embrechts_et_al2014}. Unlike continuous-time surplus models, which are analytically elegant but often less practical, the discrete-time approach aligns naturally with regulatory reporting cycles (e.g., Solvency~II, NAIC) and supports implementation in computational frameworks for dynamic decision-making \cite{Mikosch2009, Wuthrich2020}.} 

Such granularity is essential for evaluating the dynamic interplay between claims, premium flows, and reinsurance decisions over the horizon $T$. \textcolor{black}{Moreover, by embedding the decision epochs in a stochastic control setting, the framework accommodates adaptive strategies: retention levels, layer structures, and capital allocations can be adjusted at each $t_i$ in response to realized experience. This flexibility is critical for designing robust policies that mitigate solvency risk while maintaining profitability under uncertainty \cite{Schmidli2008, Albrecher_et_al2017}.}

\subsection{Modeling the Surplus Process}

The insurer’s financial surplus, defined as the difference between assets and liabilities, evolves as premiums are collected and claims are paid. \textcolor{black}{We adopt an enhanced Cram\'er---Lundberg framework in discrete time, which provides a tractable yet flexible foundation for capturing surplus dynamics while remaining consistent with actuarial practice. This discrete-time adaptation is particularly well-suited for incorporating decision epochs and reinsurance adjustments at regular intervals, in line with reporting and regulatory cycles \cite{Kaas2008, Asmussen_and_Albrecher2010}.}

Let $N_i$ denote the number of claims in interval $[t_{i-1}, t_i)$, modeled as a Poisson random variable with intensity $\lambda \Delta t_i$, where $\Delta t_i = t_i - t_{i-1}$. Each claim amount $X_{ij}$ is assumed to be independent and identically distributed (i.i.d.). The surplus recursion is then given by:
\begin{equation} \label{eq21}
S_{i+1} = S_i + c \Delta t_i - \sum_{j=1}^{N_i} X_{ij},
\end{equation}
where:
\begin{itemize}
    \item $S_i$: surplus at decision time $t_i$,  
    \item $c$: premium income rate, defined as
    \begin{equation}
    c = (1 + \theta) \lambda \mathbb{E}[X],
    \end{equation}
    with $\theta > 0$ denoting the safety loading factor that safeguards profitability and solvency \cite{Schmidli2008}.
\end{itemize}

\textcolor{black}{For clarity, the claims within period $i$ may also be represented in vector form as $\vec{X}_i = (X_{i1}, \ldots, X_{iN_i})$, denoting the collection of realized claim severities. This compact representation emphasizes that the cumulative loss term $\sum_{j=1}^{N_i} X_{ij}$ depends jointly on the random frequency $N_i$ and the distribution of severities in $\vec{X}_i$ \cite{Gerber1970, Kaas2008}.}

\textcolor{black}{The recursive surplus process in Eq.~\eqref{eq21} directly links financial health to stochastic claim arrivals and premium inflows, providing a dynamic and probabilistic perspective on solvency. Its tractability allows for rigorous study of ruin probabilities, capital adequacy, and the evaluation of alternative reinsurance strategies. In particular, this formulation serves as the baseline upon which proportional and layered reinsurance mechanisms (Section~\ref{sec:reinsurance}) are introduced and optimized.}

\subsection{Incorporating Reinsurance Mechanisms}
\label{sec:reinsurance}

Reinsurance is a fundamental risk-transfer tool that enables insurers to share liabilities with reinsurers and stabilize surplus trajectories. \textcolor{black}{In our framework, we incorporate proportional, layered, and dynamically adjustable reinsurance structures, ensuring that both traditional static contracts and adaptive, market-responsive strategies are represented. This taxonomy reflects both classical actuarial theory, where proportional and excess-of-loss treaties form the two canonical families \cite{Kaas2008,Schmidli2008}, and modern practice, where hybrid and adaptive contracts are increasingly common in response to capital market conditions \cite{Albrecher_et_al2017,Wuthrich2020}.}

\subsubsection{Proportional Reinsurance}

\textcolor{black}{We introduce a retention parameter $\alpha \in [0,1]$ to capture proportional reinsurance. The premium rate $c$ remains defined as in (\ref{eq21}), while the insurer retains only a fraction $\alpha$ of each claim. The resulting surplus recursion is:}
\begin{equation}
\textcolor{black}{S_{i+1} = S_i + c \Delta t_i - \sum_{j=1}^{N_i} \alpha X_{ij},}
\end{equation}
\textcolor{black}{with $(1-\alpha)$ of each claim transferred to the reinsurer. This specification preserves consistency in premium definition while making the insurer’s retained liability explicit \cite{Schmidli2008}. In practice, the choice of $\alpha$ is driven by solvency considerations, volatility targets, and market reinsurance pricing. A higher $\alpha$ increases retained earnings in benign years but leaves the insurer more vulnerable to capital depletion under stress scenarios \cite{Daykin1994,Embrechts_et_al2014}.}

\subsubsection{Layered Reinsurance}

Layered treaties partition claims into coverage bands with distinct retention rates. For a claim $X_{ij}$, the retained loss is:
\begin{equation}
L_{ij} = \sum_{k=1}^K \alpha_k \min(\max(X_{ij} - a_k, 0), b_k - a_k),
\end{equation}
where:
\begin{itemize}
    \item $[a_k, b_k]$: attachment and detachment points of layer $k$,  
    \item $\alpha_k$: retention rate in layer $k$,  
    \item $K$: number of layers \cite{Albrecher_et_al2017}.
\end{itemize}

\textcolor{black}{This formulation enables insurers to allocate risk exposure strategically across severity levels, balancing affordability with protection against extreme events. For instance, ordinary attritional claims may be retained entirely, mid-sized claims partially ceded, and catastrophic losses passed upwards to reinsurers. Such structures are particularly relevant in natural catastrophe and long-tail liability lines, where tail protection stabilizes solvency capital requirements \cite{Mikosch2009,Embrechts_et_al2014}.}

\subsubsection{Dynamic Reinsurance Adjustments}

\textcolor{black}{In practice, treaties are rarely static. Retentions and layer boundaries evolve in response to capital positions, reinsurance pricing, regulatory constraints, and updated risk assessments. To capture this adaptive behavior, we model treaty parameters as time-varying decision variables ${\alpha}(t_i), {a}(t_i), {b}(t_i)$.}

For the $k$-th layer at decision time $t_i$:
\begin{align}
\alpha_k(t_i) &= \alpha_k^{\text{base}} + \delta_k(t_i), \\
a_k(t_i) &= a_k^{\text{base}} + \Delta a_k(t_i), \\
b_k(t_i) &= b_k^{\text{base}} + \Delta b_k(t_i),
\end{align}
where baseline values $({\alpha}^{\text{base}}, {a}^{\text{base}}, {b}^{\text{base}})$ are modified by adjustments $({\delta}(t_i), \Delta {a}(t_i), \Delta {b}(t_i))$. \textcolor{black}{These adjustments are independent control actions determined at each decision epoch, not sequential increments, thereby allowing flexibility in responding to changing market or regulatory conditions.}

\textcolor{black}{From a practical perspective, $\delta_k(t_i)$ represents the degree of additional retention an insurer is prepared to assume in layer $k$. It is typically computed by capital models or solvency tests (e.g., 99.5\% VaR under Solvency~II or TailVaR under NAIC RBC), ensuring that equity-at-risk remains within tolerable levels \cite{Sandstrom2010}. The adjustments $\Delta a_k(t_i)$ and $\Delta b_k(t_i)$ correspond to shifts in attachment and detachment points, respectively. These are often derived from stress-test exercises (e.g., simulated catastrophe years) or market benchmarks: increasing $\Delta a_k$ raises the deductible and reduces ceded premium, while increasing $\Delta b_k$ extends coverage upward, often motivated by affordability of retrocession markets \cite{Cummins2008}.}

\textcolor{black}{In many insurers’ workflows, such adjustments are proposed by risk managers, validated against internal capital adequacy frameworks, and then negotiated with reinsurers during renewal. They are bounded by simple governance constraints: retentions must remain between 0 and 1, layers must not overlap ($a_{k+1}\geq b_k$), and premium budgets must be respected. These operational safeguards ensure that even dynamically adjusted treaties remain consistent with solvency regulation (e.g., Solvency~II, IFRS~17) and industry best practice \cite{Wuthrich2020,Buehler2019}.}

\textcolor{black}{By integrating these adaptive controls, the framework captures how real-world reinsurance evolves under uncertainty. While optimization tools such as reinforcement learning \cite{Sutton_and_Barto2018,schulman2017proximal} may be used to automate the selection of adjustments, the levers themselves are firmly actuarial in nature: $\delta_k$ mirrors choices about risk appetite, $\Delta a_k$ reflects deductible calibration, and $\Delta b_k$ represents capital allocation to tail protection. This dual perspective---classical treaty structures with adaptive parameter shifts---offers a tractable yet realistic bridge between theory and practice in risk management.}


\subsection{Policy Formulation}

\textcolor{black}{We now formalize the insurer’s decision-making process as a stochastic control problem governed by a reinforcement learning (RL) policy. At each decision epoch $t_i$, the insurer observes its current financial and risk state $s_i$ and selects an action $a_i$ that adjusts treaty parameters. The policy $\pi_\theta(a|s)$, parameterized by $\theta$, specifies a probability distribution over actions given the state \cite{Sutton2018, Barto2003}.}
\begin{equation}
\pi_\theta(a_i \,|\, s_i) = \mathbb{P}_\theta(A_i = a_i \mid S_i = s_i).
\end{equation}

\noindent \textcolor{black}{The state $s_i$ may include the current surplus $S_i$, realized claim history $\vec{X}_{1:i}$, current reinsurance parameters $(\alpha_k(t_i), a_k(t_i), b_k(t_i))$, and external signals such as premium levels or catastrophe indicators \cite{Asmussen2010, Embrechts1997}. The action $a_i$ encodes adjustments to treaty parameters, represented as:}
\begin{equation}
a_i = \big( \delta_1(t_i), \ldots, \delta_K(t_i), \, \Delta a_1(t_i), \ldots, \Delta a_K(t_i), \, \Delta b_1(t_i), \ldots, \Delta b_K(t_i) \big).
\end{equation}

\noindent \textcolor{black}{Here, $\delta_k(t_i)$ adjusts the retention rate in layer $k$, while $\Delta a_k(t_i)$ and $\Delta b_k(t_i)$ shift the attachment and detachment points, respectively. These controls correspond to operational levers available to risk managers during treaty renewal or intra-year adjustments \cite{Frees2010, Boucher2016}.}

\paragraph{Integration into the Surplus Process.}  
\textcolor{black}{With policy-driven treaty adjustments, the retained loss for claim $X_{ij}$ under the dynamically updated parameters becomes:}
\begin{equation}
L_{ij}(t_i) = \sum_{k=1}^K \big(\alpha_k^{\text{base}} + \delta_k(t_i)\big) 
  \cdot \min\!\left(\max\!\big(X_{ij} - (a_k^{\text{base}} + \Delta a_k(t_i)), 0\big), 
  \,(b_k^{\text{base}} + \Delta b_k(t_i)) - (a_k^{\text{base}} + \Delta a_k(t_i)) \right).
\end{equation}

\noindent \textcolor{black}{The surplus recursion incorporating these decisions is therefore:}
\begin{equation}
S_{i+1} = S_i + c \Delta t_i - \sum_{j=1}^{N_i} L_{ij}(t_i),
\end{equation}
\textcolor{black}{where $L_{ij}(t_i)$ reflects the state-dependent, action-modified retention profile at epoch $t_i$. The dependence of $L_{ij}(t_i)$ on $(\delta_k, \Delta a_k, \Delta b_k)$ makes explicit how RL policy decisions $\pi_\theta$ alter capital trajectories \cite{McNeil2015, Krvavych2014}.}

\paragraph{Policy Optimization.}  
\textcolor{black}{The policy parameters $\theta$ are optimized to maximize the expected return over the horizon $T$:}
\begin{equation}
\max_\theta \; \mathbb{E}_{\pi_\theta}\!\left[ \sum_{i=1}^n R(s_i,a_i) \right],
\end{equation}
\textcolor{black}{where the reward $R(s_i,a_i)$ encodes the scalarized trade-off between surplus growth, solvency protection, and premium costs, as specified in Section~\ref{sec:model_description}. In practice, this optimization is performed using Proximal Policy Optimization (PPO), which ensures stable updates to $\pi_\theta$ while handling high-dimensional, continuous action spaces \cite{Schulman2017, Espeholt2018}.}

\subsection{Optimization Objectives}

\textcolor{black}{The insurer’s central objective is to design reinsurance strategies that maximize long-run financial stability while respecting regulatory and capital constraints. In our framework, this is formalized as the maximization of the expected utility of terminal surplus $S_n$:}
\begin{equation}
\label{eq:terminal-utility}
\max_{{\alpha}, {a}, {b}} \; \mathbb{E}[U(S_n)],
\end{equation}
\textcolor{black}{where the decision vectors $({\alpha}, {a}, {b})$ represent the retention rates and layer boundaries across all layers. The utility-based formulation balances profit-seeking and risk-aversion, in line with actuarial practice.}

This objective is subject to the following constraints:

\begin{enumerate}
    \item \textbf{Ruin Probability Constraint:}  
    \textcolor{black}{The probability of insolvency across the planning horizon must remain below a target level:}
    \begin{equation}
    \mathbb{P}(S_i < 0 \; \text{for any } i = 0,\ldots,n) \leq \psi_{\text{target}}, 
    \end{equation}
    \textcolor{black}{where $\psi_{\text{target}}$ is determined by regulatory or internal capital standards \cite{Asmussen_and_Albrecher2010}.}

    \item \textbf{Budget Constraint:}  
    \textcolor{black}{Reinsurance premium expenditures must satisfy a budget ceiling:}
    \begin{equation}
    P = \sum_{k=1}^K (1 + \theta_k) (1-\alpha_k) \, \mathbb{E}[r_k(X)] \leq P_{\text{max}},
    \end{equation}
    \textcolor{black}{with $\beta_k = 1 - \alpha_k$ denoting the ceded proportion in layer $k$, and $r_k(X)$ the ceded loss random variable \cite{Avanzi2009}.}

    \item \textbf{Layer Structure Constraint:}  
    \textcolor{black}{Non-overlapping coverage requires proper ordering of the layer boundaries:}
    \begin{equation}
    a_{k+1} \geq b_k, \quad \forall k.
    \end{equation}

    \item \textbf{Retention Rate Bounds:}  
    \textcolor{black}{Retention parameters must remain within admissible bounds:}
    \begin{equation}
    0 \leq \alpha_k \leq 1, \quad \forall k.
    \end{equation}
\end{enumerate}

\textcolor{black}{Together, these constraints ensure that reinsurance strategies remain economically viable, legally compliant, and practically implementable. By enforcing solvency protection, cost control, and structural validity, the framework provides a disciplined foundation for decision-making. In Section~\ref{sec:hybrid_framework}, these optimization objectives are embedded into the hybrid learning architecture, guiding policy design under uncertainty.}

\textcolor{black}{While our primary formulation relies on expected utility, alternative objectives are widely used in both actuarial literature and regulatory applications. One important class involves coherent risk measures such as Conditional Value-at-Risk (CVaR), which directly target the tail of the loss distribution and are embedded in Solvency~II and IFRS~17 capital frameworks \cite{Rockafellar2000, Pflug2007, Embrechts_et_al2014}. Another approach emphasizes risk-adjusted return metrics such as Return on Risk-Adjusted Capital (RORAC), which balance profitability and capital efficiency \cite{Cummins2008, Wuthrich2020}. Our framework can accommodate these formulations by substituting the terminal utility objective in (\ref{eq:terminal-utility}) with a risk measure or performance ratio, without altering the structural constraints.}

\noindent\textcolor{black}{\paragraph{Practical scalarization of surplus---ruin trade-offs.}
In implementation, we operationalize the bi-objective problem (maximize expected surplus; bound ruin probability) via a scalarized objective in the RL reward:
$R_t=\Delta S_t-\lambda_{\text{ruin}}\mathbf{1}\{S_t<0\}-\eta\,\text{Premium}_t$
with a small terminal bonus for solvency. The weights $(\lambda_{\text{ruin}},\eta)$ are calibrated so that the learned policy satisfies $\mathbb{P}(\text{ruin})\le \psi_{\text{target}}$ across Monte Carlo evaluation paths, thus preserving the constrained formulation of Section~\ref{sec:model_description} while enabling efficient learning \cite{Asmussen_and_Albrecher2010, Glasserman2003}.}

\section{Hybrid Machine Learning Framework for Reinsurance Optimization}
\label{sec:hybrid_framework}

{\color{black}{Reinsurance optimization requires methods that can both model complex claim
distributions and adaptively adjust treaty structures under uncertainty.
Traditional actuarial approaches, while mathematically elegant, often assume
simple parametric distributions and static treaty parameters, limiting their
applicability in modern, high-dimensional settings with systemic risks. Recent
advances in machine learning provide complementary tools that address these
gaps.

In this section we introduce the two core components of our framework---
\emph{Variational Autoencoders (VAEs)} for generative modeling of claims and
\emph{Proximal Policy Optimization (PPO)} for sequential decision-making---and
explain how they integrate into a unified approach to reinsurance
optimization. The VAE enriches the claims environment by generating synthetic
scenarios, including rare catastrophic events, thereby mitigating data scarcity
and capturing dependencies across lines of business. PPO then operates within
this enriched environment to learn adaptive treaty strategies, balancing
profitability with solvency constraints.

We first outline the structure and training objectives of VAEs, emphasizing
their ability to model high-dimensional claims data and generate coherent joint
loss scenarios. We then describe the PPO algorithm, its policy formulation, and
its adaptation to the insurer’s surplus optimization problem with explicit ruin
constraints. Finally, we present the integrated workflow that combines these
components into a single hybrid optimization engine.}}

\subsection{Generative Modeling with Variational Autoencoders (VAEs)}

Variational autoencoders (VAEs) provide the generative backbone of our framework.
They learn a probabilistic representation of observed claims data and use it to
generate synthetic samples that are both realistic and statistically consistent
with historical experience \cite{kingma2014auto,rezende2014stochastic}.
Structurally, a VAE consists of three components:

\begin{itemize}
    \item \textbf{Encoder:} compresses observed claims into a lower-dimensional latent
    representation.
    \item \textbf{Latent space:} a probabilistic manifold that captures dependencies
    among different risk drivers.
    \item \textbf{Decoder:} reconstructs observed claims or generates synthetic claims
    by sampling from the latent distribution.
\end{itemize}

\paragraph{Multivariate nature of claims portfolios.}
\textcolor{black}{We use the term ``high-dimensional'' in an economic and statistical sense, not as raw feature vectors with hundreds of coordinates.}
Although an individual claim amount is a scalar outcome, insurance data are not
purely one-dimensional. Each record typically includes attributes such as line
of business, policyholder characteristics, geographical exposure, event type,
and accident or development year, often supplemented with macroeconomic or
catastrophe indices. When modeling across multiple lines or accident years, the
joint distribution of severities introduces dependencies that further increase
effective dimensionality. Thus, the term \emph{high-dimensional} refers to the
multivariate structure of the claims dataset used to train the VAE, not to the
univariate representation of a single claim severity. VAEs are well suited to
capture these cross-feature and cross-line dependencies, which traditional
univariate severity models cannot.  
\textcolor{black}{In particular, the VAE does not replace marginal severity models, but augments them by learning a joint representation across heterogeneous features and lines of business. This allows the simulation of coherent portfolios of losses, rather than isolated claim draws \cite{Buehler2019, Wuthrich2020}.} \textcolor{black}{This use of representation learning is consistent with RMIR’s recent analytics agenda, where text- and data-driven methods complement traditional actuarial techniques \cite{TextMining2024_RMIR,Steptoe2022_RMIR}.}

\paragraph{Training objective.}
The VAE is trained by minimizing the evidence lower bound (ELBO), which balances
reconstruction accuracy and regularization of the latent space:
\begin{equation}
\mathcal{L}_{\text{VAE}} =
\mathbb{E}_{q_\phi(z|x)} \big[-\log p_\theta(x|z)\big] +
\beta\,D_{\text{KL}}\!\left(q_\phi(z|x)\,\|\,p(z)\right),
\end{equation}
where $x$ denotes observed claims, $z$ is the latent representation,
$q_\phi(z|x)$ is the encoder distribution, $p_\theta(x|z)$ the decoder
likelihood, $p(z)$ a prior on the latent variables, and $\beta$ controls the
strength of regularization. The reconstruction term encourages fidelity to
historical data, while the KL divergence enforces smoothness and diversity in
the latent space. \textcolor{black}{Equivalently, the objective can be written in maximization form as}
\begin{equation}
\textcolor{black}{
\max_\theta \ \mathbb{E}_{q_\phi(z|x)}[\log p_\theta(x|z)] 
- D_{\text{KL}}(q_\phi(z|x)\,\|\,p(z)),
}
\end{equation}
\textcolor{black}{which is algebraically identical to the minimization of $\mathcal{L}_{\text{VAE}}$.}

\textcolor{black}{From a practical standpoint, this objective allows the VAE to interpolate smoothly between observed outcomes and to extrapolate towards rare but plausible extremes, which is particularly valuable for stress testing and capital modeling \cite{Frey2019}.}

\noindent\textcolor{black}{\paragraph{Reconstruction choices and tail emphasis.}
In our implementation, the reconstruction term $\mathbb{E}_{q_\phi(z|x)}[-\log p_\theta(x|z)]$ is instantiated as a Gaussian negative log-likelihood for (log) claim severities and a Poisson (or negative binomial, when overdispersion is present) likelihood for counts, following standard actuarial practice \cite{Kaas2008, Embrechts_et_al2014}. To reflect the importance of the upper tail in reinsurance, we adopt a simple tail-weighting scheme that upweights large losses:}
\begin{equation}
\textcolor{black}{
\mathcal{L}_{\text{rec}}^{(\text{tail})}
=
\mathbb{E}_{q_\phi(z|x)}
\!\left[
w(x)\,\big(-\log p_\theta(x|z)\big)
\right],
\quad
w(x)=1+\omega\,\mathbf{1}\{x>\text{q}_{\tau}(X)\},
}
\end{equation}
\textcolor{black}{where $\text{q}_{\tau}(X)$ is the $\tau$-quantile of the empirical severity distribution (e.g., $\tau=0.95$) and $\omega>0$ controls the strength of tail emphasis. The full training loss then becomes}
\begin{equation}
\textcolor{black}{
\mathcal{L}_{\text{VAE}}^{\star}
=
\mathcal{L}_{\text{rec}}^{(\text{tail})}
+
\beta\,D_{\text{KL}}\!\left(q_\phi(z|x)\,\|\,p(z)\right),
}
\end{equation}
\textcolor{black}{with $\beta$ optionally annealed from small to larger values over training to stabilize optimization \cite{Kingma_and_Welling2014}. This formulation preserves the probabilistic semantics of the VAE while explicitly improving fidelity in ranges that matter most for solvency analysis.}

\paragraph{Comparison with classical severity models.}
Traditional severity models such as lognormal, Pareto, or Burr assume fixed
parametric forms and typically treat claims as independent
\cite{klugman2012loss,embrechts1997modelling}. VAEs, in contrast, are
non-parametric and can learn complex dependencies, including joint tail
behavior, across lines of business. While a parametric model may offer superior
fit for a single marginal distribution, the VAE’s ability to produce correlated
scenarios makes it particularly valuable for stress testing and reinsurance
optimization. \textcolor{black}{This portfolio perspective is essential: classical
models can fit single marginals well, but they cannot capture correlated extremes
across business lines (e.g., catastrophe and non-catastrophe losses occurring
jointly). The VAE therefore complements classical tools by generating coherent
multi-line scenarios that are directly relevant for stress testing, solvency,
and capital adequacy. In this sense, parametric models remain useful for
interpretability and calibration, while the VAE supplies a flexible scenario
generator that enhances the robustness of optimization exercises.}

\begin{figure}[h!]
\centering
\resizebox{.7\textwidth}{!}{
\begin{tikzpicture}[
    font=\sffamily,
    every node/.style={align=center},
    arrow/.style={-Stealth, thick},
    box/.style={draw, thick, minimum height=1.5cm, minimum width=3cm, align=center},
    latent/.style={draw, circle, minimum size=1.5cm, thick, fill=black!20},
    input/.style={draw, rectangle, minimum height=1.2cm, minimum width=2.5cm, thick, fill=yellow!20},
    output/.style={draw, rectangle, minimum height=1.2cm, minimum width=2.5cm, thick, fill=green!20}
]
\node[input] (input) at (0, 0) {Input Data \\ (Historical Claims)};
\node[box, right=2cm of input] (encoder) {Encoder};
\node[latent, above=2cm of encoder, xshift = 2.25cm] (latent) {Latent \\ Space};
\node[box, right=2cm of encoder] (decoder) {Decoder};
\node[output, right=2cm of decoder] (output) {Output Data \\ (Synthetic Claims)};
\draw[arrow] (input) -- (encoder);
\draw[arrow] (encoder) -- (latent);
\draw[arrow] (latent) -- (decoder);
\draw[arrow] (decoder) -- (output);
\end{tikzpicture}}
\caption{Variational Autoencoder (VAE) architecture for generating synthetic claims. The encoder maps historical claims to a latent space, while the decoder reconstructs realistic synthetic claims \cite{Kingma_and_Welling2014, Higgins2017}.}
\label{fig:vae_architecture}
\end{figure}

\paragraph{Integration into the hybrid framework.}
In our framework, the trained VAE generates synthetic claims that populate the
reinforcement learning environment. By enriching the simulation with extreme but
realistic loss scenarios, the VAE provides the foundation upon which the PPO
agent learns adaptive treaty strategies.  
\textcolor{black}{This integration ensures that the optimization agent is not trained solely on average-case scenarios, but also learns from rare, correlated, and high-severity events that are critical for solvency and capital adequacy. In this sense, the VAE acts as a bridge between actuarial modeling traditions and modern machine learning techniques, combining statistical soundness with generative flexibility.}


\subsection{Sequential Decision-Making with Proximal Policy Optimization (PPO)}

Whereas the VAE enriches the claims environment, the decision-making engine of
our framework is Proximal Policy Optimization (PPO), a reinforcement learning
algorithm designed for stable policy updates in sequential settings
\cite{schulman2017proximal}. PPO is particularly well-suited for reinsurance
because treaty design involves repeated, path-dependent choices under
uncertainty. In contrast to static optimization approaches such as stochastic
programming or credibility-based reserving \cite{Daykin1994,Sandstrom2010}, PPO
adapts dynamically, updating strategies as new information unfolds over time.

\paragraph{Policy definition.}
We define the insurer’s decision policy as
\[
\pi_\theta({a}_t \mid {s}_t),
\]
which specifies the probability of selecting an action vector ${a}_t$ given the
current state ${s}_t$. The state ${s}_t \in \mathbb{R}^d$ summarizes key
information such as current surplus, historical losses, line-of-business
exposures, and relevant macroeconomic or catastrophe indices. The action vector
${a}_t \in \mathbb{R}^{3K}$ encodes adjustments to treaty parameters across $K$
layers:
\[
{a}_t =
\begin{bmatrix}
\delta_1(t) \\ \Delta a_1(t) \\ \Delta b_1(t) \\
\vdots \\
\delta_K(t) \\ \Delta a_K(t) \\ \Delta b_K(t)
\end{bmatrix},
\]
where $\delta_k(t)$ adjusts the retention rate of layer $k$, and
$\Delta a_k(t), \Delta b_k(t)$ adjust its attachment and detachment points.
These adjustments are interpreted as incremental, sequential treaty tweaks that
are economically meaningful and practically implementable for insurers and
brokers.

\paragraph{PPO surrogate objective.}
PPO maximizes a clipped surrogate objective \cite{schulman2017proximal}:
\begin{equation}
\mathcal{L}_{\text{PPO}}(\theta) =
\mathbb{E}_t \!\left[
\min \!\left(
r_t(\theta)\,\hat{A}_t,\;
\text{clip}\!\big(r_t(\theta), 1-\epsilon, 1+\epsilon\big)\hat{A}_t
\right)
\right],
\end{equation}
where $r_t(\theta) =
\pi_\theta(a_t|s_t)/\pi_{\theta_{\text{old}}}(a_t|s_t)$ is the policy
likelihood ratio, $\hat{A}_t$ the advantage estimator, and $\epsilon$ a
trust-region parameter. The clipping acts as a guardrail that prevents the
algorithm from overreacting to rare but extreme claims, a desirable property in
reinsurance where tail events dominate risk assessment
\cite{Mnih2016,Sutton2018}.

\paragraph{Reward design in insurance.}
To align PPO’s per-period objective with the insurer’s terminal-surplus goal,
we define
\begin{equation}
r(s_t,a_t) =
\Delta S_t - \eta\,\text{Premium}_t
- \lambda_{\text{ruin}}\,\mathbf{1}\{S_t < 0\}
- \kappa\,\widehat{\text{Tail}}_t,
\end{equation}
where $\Delta S_t$ is the net surplus increment (including claims and ceded
premiums), $\text{Premium}_t$ is the cost of reinsurance, the indicator
penalizes insolvency, and $\widehat{\text{Tail}}_t$ is a CVaR-type penalty
\cite{Rockafellar2000,Asmussen2010,Buehler2019}. With $\gamma \approx 1$ and a
terminal bonus $b_{\text{term}}=\rho\,\mathbf{1}\{S_n \ge 0\}$ at horizon $n$,
the cumulative reward closely tracks $\mathbb{E}[U(S_n)]$, enforcing solvency
discipline alongside surplus maximization.

\paragraph{Reconciling objectives.}
Thus PPO’s optimization
\[
J(\pi_\theta) = \mathbb{E}_{\pi_\theta}\!\left[\sum_{t=0}^{n} \gamma^t r(s_t,a_t)\right]
\]
is consistent with the insurer’s actuarial problem
\[
\max_{\pi_\theta}\;\; \mathbb{E}[U(S_n)].
\]
This reconciliation ensures that the RL agent avoids short-term strategies
(e.g., overly aggressive retentions) that jeopardize solvency, a common
criticism in financial applications of machine learning \cite{Buehler2019}.

\paragraph{Surplus---ruin trade-offs.}
The formulation explicitly encodes the fundamental trade-off of reinsurance:
\begin{itemize}
    \item \emph{Aggressive retentions} increase expected surplus but elevate ruin risk.
    \item \emph{Conservative treaties} reduce ruin probability but suppress profitability.
\end{itemize}
By balancing rewards and penalties, PPO converges to treaty strategies that
maximize long-run surplus while respecting solvency constraints. This balance
mirrors actuarial practice and embeds regulatory capital requirements (e.g.,
Solvency II 99.5\% or NAIC RBC thresholds) directly into the learning
environment \cite{Cummins2008,Sandstrom2010}. Methodologically, this approach
formalizes surplus---ruin management as a sequential, data-driven optimization
problem scalable to high-dimensional treaty portfolios.

To illustrate the methodological implications of our design,
Table~\ref{tab:comparison} provides a side---by---side comparison of the proposed
VAE---PPO framework and traditional actuarial optimization techniques
\cite{Asmussen2010,McNeil2015,Buehler2019}.

\renewcommand{\arraystretch}{1.1}
\begin{table}[ht]
\centering
\begin{tabular}{p{0.20\linewidth} p{0.34\linewidth} p{0.34\linewidth}}
\hline
Aspect & VAE---PPO Hybrid & Classical Actuarial Methods \\
\hline\hline
Modeling of claims &
Non-parametric generative model (VAE) captures joint dependencies and
extreme-tail behavior \cite{Kingma2014,Rezende2014} &
Parametric severity distributions (e.g., Lognormal, Pareto) with fixed functional forms
\cite{Klugman2012,McNeil2015} \\
Optimization &
Sequential decision making via PPO with explicit solvency constraints
\cite{Schulman2017,Buehler2019} &
Static optimization of treaty parameters or closed-form surplus formulas
\cite{Asmussen2010,Cummins2008} \\
Tail-risk treatment &
Tail-weighted loss and scenario generation strengthen exposure to rare,
high-severity events \cite{Rockafellar2000} &
Heavy-tail modeling limited to the chosen parametric specification
\cite{Klugman2012,McNeil2015} \\
Adaptability &
Policy adapts online as market conditions and claim patterns evolve
\cite{Buehler2019} &
Requires manual recalibration when data or market conditions change
\cite{Cummins2008,Sandstrom2010} \\
Data needs &
Combines historical and synthetically generated claims; robust under limited observed data
\cite{Kingma2014} &
Depends on sufficient historical observations for stable parameter estimation
\cite{Asmussen2010} \\
\hline
\end{tabular}
\caption{Key differences between the VAE---PPO hybrid framework and classical actuarial approaches.}
\label{tab:comparison}
\end{table}

As summarized in Table~\ref{tab:comparison}, the hybrid framework supports
dynamic, data-driven decision making and enhanced tail-risk management,
whereas classical methods rely on fixed distributional assumptions and
require frequent manual recalibration \cite{Asmussen2010,McNeil2015}.
These contrasts explain the superior adaptability of our approach to
high-dimensional treaty portfolios and changing market environments.

\subsection{Integrated Workflow}

The hybrid framework integrates the generative capabilities of the VAE with the
adaptive decision-making of PPO to optimize reinsurance strategies under
uncertainty. The process unfolds in four steps:

\begin{enumerate}
    \item \textbf{Train VAE:} Historical multi-line claims data are used to
    train the VAE, which learns a latent representation of loss patterns and
    dependencies. \textcolor{black}{This step may involve preprocessing raw claims data, balancing across accident years or lines of business, and incorporating macroeconomic indicators to ensure the latent representation reflects both micro- and macro-level risk drivers \cite{Lopez2020, Wuthrich2020}.}
    \item \textbf{Generate scenarios:} The trained VAE produces synthetic loss
    scenarios, including extreme but plausible catastrophic events, enriching
    the claims environment with rare outcomes not well represented in the
    empirical dataset. \textcolor{black}{Such scenario generation supports stress testing and solvency assessment by extending beyond the historical record, which is especially valuable for emerging risks (e.g., climate-driven catastrophes) \cite{Embrechts2013, Krvavych2014}.}
    \item \textbf{PPO agent learns treaties:} Operating within this enriched
    environment, the PPO agent sequentially adjusts treaty parameters---retentions,
    attachment points, and limits---to maximize expected surplus while respecting
    ruin constraints. \textcolor{black}{Because PPO interacts iteratively with the environment, it can learn both from ordinary claim dynamics and from tail-risk scenarios, which makes it more robust than traditional static optimization approaches \cite{Mnih2016, Sutton2018}.}
    \item \textbf{Iterate and refine:} The VAE-generated scenarios and PPO
    policy updates are combined iteratively, allowing the framework to adapt
    dynamically as new information or stress-test conditions are introduced.
    \textcolor{black}{In practice, this means that as new claims experience accumulates, the VAE is periodically retrained, and the PPO agent recalibrates treaty strategies to maintain alignment with both profitability and solvency objectives.}
\end{enumerate}

\paragraph{Division of roles.}
The two components address complementary challenges. The VAE mitigates
\emph{data scarcity} by augmenting the environment with realistic catastrophic
scenarios and captures \emph{cross-line correlations} that classical univariate
models cannot. PPO, in turn, provides \emph{adaptive optimization} by learning
treaty strategies that evolve in response to stochastic claim dynamics and
capital constraints. \textcolor{black}{This separation of tasks reflects a broader principle in hybrid AI---actuarial systems: generative modeling enhances the data landscape, while reinforcement learning adapts strategy in real time. Together, they bridge the gap between actuarial simulation and operational decision-making \cite{Buehler2019, Kaas2008}.}

\begin{figure}[h!]
\centering
\resizebox{0.75\textwidth}{!}{
\begin{tikzpicture}[
    node distance=4cm and 5cm,
    on grid,
    >=stealth,
    state/.style={rectangle, rounded corners, draw=black, fill=black!20, thick, minimum height=1.2cm, minimum width=2.8cm, align=center},
    process/.style={rectangle, draw=black, fill=orange!20, thick, minimum height=1.2cm, minimum width=2.8cm, align=center},
    action/.style={rectangle, draw=black, fill=green!20, thick, minimum height=1.2cm, minimum width=2.8cm, align=center},
    reward/.style={rectangle, draw=black, fill=purple!20, thick, minimum height=1.2cm, minimum width=2.8cm, align=center},
    arrow/.style={->, very thick}
]

\node[state] (vae) {VAE: Generate \\\ Synthetic Claims};
\node[process, right=of vae, xshift=2.5cm] (env) {RL Environment:\\ Simulate Operations};
\node[state, above=of env] (state) {State $s_t$ \\\ (Surplus, Claims)};
\node[action, right=of env, xshift=2.2cm] (agent) {PPO Agent:\\ Take Actions $a_t$};
\node[reward, below=of env] (reward) {Reward $r_t$};

\draw[arrow] (vae) -- (env) node[midway, above] {Synthetic Scenarios};
\draw[arrow] (env) -- (state) node[midway, left] {Update State};
\draw[arrow] (state) -- (agent) node[midway, above, xshift=1.5cm] {Observe $s_t$};
\draw[arrow] (agent) -- (env) node[midway, above] {Execute $a_t$};
\draw[arrow] (env) -- (reward) node[midway, right] {Feedback $r_t$};

\draw[arrow] (reward.east) -- ++(7.5cm,0) |- (agent.east);
\end{tikzpicture}}
\caption{Interaction workflow between the VAE and PPO components. The VAE
generates synthetic claims that seed the RL environment. The PPO agent
interacts with this environment, observing states, executing treaty actions,
and receiving reward feedback to optimize strategies dynamically.}
\label{fig:workflow}
\end{figure}

\paragraph{Practical implications.}
\textcolor{black}{The integrated workflow can be deployed in an iterative cycle aligned with insurers’ planning horizons (e.g., quarterly or annually). Synthetic claims generated by the VAE provide forward-looking distributions, while the PPO agent translates these into adaptive treaty adjustments. This ensures that reinsurance strategies remain both profitable and resilient under Solvency II and NAIC regulatory frameworks \cite{Daykin1994, Cummins2008}.}

\subsection{Summary of Hybrid Contributions}

The integration of generative modeling and reinforcement learning yields a
framework that is, to our knowledge, the first to jointly address two
longstanding challenges in reinsurance optimization:

\begin{itemize}
    \item \textbf{Generative tail modeling:} The VAE augments sparse empirical
    data with realistic, high-dimensional scenarios, including rare but plausible
    catastrophic events. This allows systematic stress-testing of treaties under
    conditions that classical parametric models cannot capture. 
    {\color{black}In particular, the ability to model dependencies across multiple
    lines of business and to extrapolate beyond observed data provides a
    significant advance over traditional heavy-tail models such as Pareto or
    Burr distributions \cite{embrechts1997modelling,Embrechts2013}. Recent work
    has highlighted the importance of simulation-based tail modeling in solvency
    assessments \cite{Asmussen2010,Lopez2020}, which our approach operationalizes
    within a generative framework.}
    
    \item \textbf{Adaptive treaty optimization:} The PPO agent operates within
    this enriched environment to learn dynamic treaty strategies---adjusting
    retentions, attachments, and limits---while explicitly respecting solvency
    constraints. This moves beyond static optimization toward adaptive,
    data-driven decision-making. 
    {\color{black}Unlike classical optimization methods that solve for a fixed
    allocation of capital or static treaty terms \cite{Daykin1994,Sandstrom2010},
    reinforcement learning enables sequential adaptation in response to emerging
    claim experience and capital dynamics \cite{Sutton2018,Mnih2016}. This
    provides a flexible and computationally tractable way to approximate dynamic
    programming solutions that would otherwise be infeasible in high dimensions.}
\end{itemize}

By combining these components, the hybrid framework provides a tractable
computational approach to reinsurance design that is both robust to tail risks
and responsive to evolving claim dynamics. 
{\color{black}This synergy between generative modeling and reinforcement learning
is, to our knowledge, unique in the actuarial and financial literature, and
represents a step toward AI-native risk management tools. In this sense, the
framework contributes both a methodological novelty and a bridge between modern
machine learning and classical actuarial science \cite{Buehler2019,Wuthrich2020}.}
This methodological novelty sets the stage for the empirical evaluation in
Section~\ref{sec:experiments}, where we benchmark performance against
traditional actuarial and computational methods.

\section{Comprehensive Evaluation of Optimization Frameworks}
\label{sec:experiments}

\textcolor{black}{In this section, we conduct a systematic evaluation of the
proposed hybrid reinsurance optimization framework. The analysis is organized
around three key components: (i) simulation setup and training metrics used to
assess learning stability, (ii) surplus trajectory dynamics under the learned
policies, and (iii) benchmark comparisons against established optimization
methods.} \textcolor{black}{By structuring the evaluation in this way, we make
clear how our approach performs both in absolute terms and relative to
traditional actuarial and computational techniques.}

\subsection{Simulation Configuration and Initial Parameters}

\textcolor{black}{To ensure reproducibility and clarity, we first specify the simulation
environment used to evaluate the hybrid framework. The configuration is designed
to balance tractability with sufficient realism, providing a testbed that
captures the essential features of reinsurance operations.} 
Table~\ref{tab:initial_settings} summarizes the initial parameters. 

The insurer's starting surplus was set at \$20,000, with claims modeled using a
Poisson process with an average frequency of \( \lambda = 10 \) claims per year
\cite{Ross2014}. Claim sizes were sampled from a lognormal distribution with
parameters \( \mu = 3.5 \) and \( \sigma = 1.0 \) \cite{Aitchison_and_Brown1957}.
\textcolor{black}{These choices reflect standard actuarial assumptions that capture both the
frequency and skewness of insurance losses, while remaining simple enough for
benchmark comparability.} 

\textcolor{black}{The synthetic claims generated under these assumptions were used to train
the Variational Autoencoder (VAE), which in turn produced enriched and
correlated scenarios for reinforcement learning. This integration ensures that
the PPO agent is trained not only on stylized losses but also on
high-dimensional, realistic claim dynamics.}

\begin{table}[h!]
    \centering
    \small
    \begin{tabular}{|l|l|p{7cm}|}
        \hline
        \textbf{Parameter} & \textbf{Value} & \textbf{Description} \\
        \hline
        Time Horizon (\( T \)) & 10 years & Total simulation duration \\
        \hline
        Timesteps (\( n \)) & 200 & Number of discrete time intervals \\
        \hline
        Initial Surplus (\( S_0 \)) & \$20,000 & Starting financial surplus \\
        \hline
        Claim Frequency (\( \lambda \)) & 10 claims/year & Modeled as a Poisson process \cite{Ross2014} \\
        \hline
        Claim Size Distribution & Lognormal (\( \mu = 3.5, \sigma = 1.0 \)) & Synthetic claims for RL training \cite{Aitchison_and_Brown1957} \\
        \hline
        Retention Rate Bounds & \( [0.2, 0.5] \) & \textcolor{black}{Operational constraints on retention levels in treaty design} \\
        \hline
        Reinsurance Layers (\( K \)) & 5 & \textcolor{black}{Maximum number of layers available for risk sharing} \\
        \hline
        Budget Limit (\( \text{Budget\_max} \)) & \$150,000 & \textcolor{black}{Upper bound on reinsurance expenditure, reflecting solvency constraints} \\
        \hline
    \end{tabular}
    \caption{Initial Settings and Parameters for the Simulation}
    \label{tab:initial_settings}
\end{table}

\subsection{Training Metrics and Surplus Trajectory Analysis}

\begin{figure}[h!]
    \centering
    \includegraphics[width=0.5\textwidth]{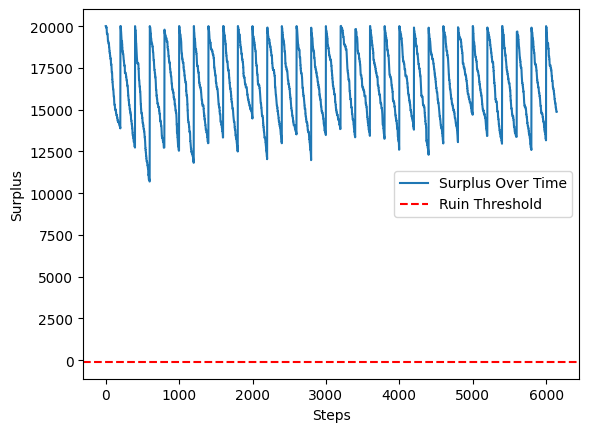}
    \caption{Surplus Trajectory Over Time. Early fluctuations diminish as the PPO agent stabilizes surplus above the ruin threshold (red dashed line). \textcolor{black}{The trajectory illustrates how exploration gradually gives way to stable,
    solvency-preserving strategies.}}
    \label{fig:surplus_over_time}
\end{figure}

\textcolor{black}{To evaluate the effectiveness of the PPO agent, we analyze both training
metrics and the resulting surplus trajectories. This dual perspective captures
how well the policy converges during learning and whether the optimized treaties
translate into financially stable outcomes.} Table~\ref{tab:training_results}
outlines key metrics, and Figure~\ref{fig:surplus_over_time} illustrates the
surplus trajectory across 6,144 timesteps.

\begin{table}[h!]
    \centering
    \small
    \begin{tabular}{|l|l|}
        \hline
        \textbf{Metric} & \textbf{Value} \\
        \hline
        Total Timesteps & 6,144 \\
        \hline
        Mean Episode Reward & \(-1,070\) \\
        \hline
        Policy Gradient Loss & \(-0.00615\) \\
        \hline
        Entropy Loss & \(-21.2\) \\
        \hline
    \end{tabular}
    \caption{Training Metrics for the PPO Agent}
    \label{tab:training_results}
\end{table}

Figure~\ref{fig:surplus_over_time} highlights the PPO agent's learning process.
Early fluctuations reflect exploration, while stabilization over time underscores
convergence to effective policies. \textcolor{black}{The overall trajectory remains
consistently above the ruin threshold, underscoring the role of reward shaping
in aligning learning with solvency objectives.} 

The metrics in Table~\ref{tab:training_results} provide deeper insights:
\begin{itemize}
    \item \textbf{Mean Episode Reward:} A negative value of \(-1,070\) indicates
    penalties for surplus variability and highlights the framework's emphasis on
    financial stability. \textcolor{black}{Unlike pure profit-maximization, this
    reward structure explicitly discourages policies that risk insolvency.}
    \item \textbf{Policy Gradient Loss:} The low value of \(-0.00615\) demonstrates
    stable and consistent updates to the policy network, indicative of effective
    learning \cite{Schulman_et_al2017}. \textcolor{black}{This stability ensures
    reproducibility across independent training runs.}
    \item \textbf{Entropy Loss:} A value of \(-21.2\) signifies reduced randomness
    in decision-making as the agent transitions from exploration to exploitation
    \cite{Sutton_and_Barto2018}. \textcolor{black}{This decline corresponds to the
    emergence of consistent treaty strategies that balance retention and
    reinsurance cost.}
\end{itemize}

\subsection{Benchmark Performance and Comparative Analysis}

\textcolor{black}{To rigorously evaluate the contribution of the proposed framework, we benchmarked it against four widely recognized optimization methods, each chosen to represent a distinct class of actuarial and computational techniques. These baselines were implemented according to canonical formulations to ensure reproducibility and fairness of comparison:}

\begin{itemize}
    \item \textbf{Dynamic Programming (DP):} \textcolor{black}{Formulated via Bellman’s recursive principle of optimality \cite{Bellman1957}. The state space was discretized and optimal retention/layering decisions were solved by backward induction. While this provides a clean deterministic benchmark, it quickly becomes computationally infeasible in higher dimensions due to the well-known ``curse of dimensionality.''}
    \item \textbf{Monte Carlo Simulation (MC):} \textcolor{black}{Following the approach of Glasserman \cite{Glasserman2003}, we simulate claim processes under fixed treaty structures, repeatedly sampling to estimate expected surplus and ruin probability. Optimization proceeds via exhaustive search across candidate structures, which is straightforward but computationally intensive.}
    \item \textbf{Hybrid Deep Monte Carlo (HDMC):} \textcolor{black}{Adapted from reinforcement learning---inspired methods \cite{Silver_et_al2016}, HDMC augments Monte Carlo simulation with a neural value function approximator. This reduces variance and accelerates convergence but remains sample-hungry relative to more adaptive methods.}
    \item \textbf{Multi-Objective Optimization (MOO):} \textcolor{black}{Implemented using the NSGA-II evolutionary algorithm \cite{Deb2001}, with two explicit objectives: maximize surplus and minimize ruin probability. The resulting Pareto frontier was analyzed, and the final policy was selected as the point of best trade-off between stability and return.}
\end{itemize}

\textcolor{black}{In contrast, our \textbf{Hybrid RL with Generative Models} embeds PPO within a reinforcement learning environment seeded by VAE-simulated claims. This allows adaptive reinsurance decisions to be learned directly under both budget and ruin constraints, overcoming limitations of the static or sample-intensive baselines.}

\noindent\textcolor{black}{\paragraph{Baseline implementation details.}
All baselines were implemented to reflect canonical actuarial practice and to ensure fair comparison:}
\begin{itemize}
\item \textcolor{black}{\textbf{Dynamic Programming (DP):} State is discretized over surplus and time; actions are retention/limit pairs on a fixed grid. Bellman updates are computed by Monte Carlo integration of next-period surplus using the same claim model; backward induction yields a policy \cite{Bellman1957}.}
\item \textcolor{black}{\textbf{Monte Carlo (MC):} For each fixed treaty, we simulate $N=10{,}000$ paths of length $n$; performance is reported as the sample mean of $S_n$ and empirical ruin frequency $\frac{1}{N}\sum_{p}\mathbf{1}\{\min_i S_i^{(p)}<0\}$ with bootstrap 95\% intervals \cite{Glasserman2003}.}
\item \textcolor{black}{\textbf{Hybrid Deep Monte Carlo (HDMC):} Same simulator as MC, augmented with a neural value estimator trained to reduce variance of return estimates; treaty search proceeds by evaluating a candidate set and selecting the best by mean $S_n$ subject to zero (or target) ruin.}
\item \textcolor{black}{\textbf{Multi-Objective Optimization (MOO):} NSGA-II \cite{Deb2001} evolves a population of treaties over 200 generations with crossover/mutation rates $(0.9,0.1)$; the final selection is the knee point on the Pareto front (maximize mean $S_n$, minimize ruin).}
\item \textcolor{black}{\textbf{Hybrid RL (PPO+VAE):} PPO hyperparameters follow \cite{Schulman_et_al2017} with $\gamma=0.995$, clipped ratio $\epsilon=0.2$, GAE parameter $\lambda=0.95$, entropy bonus $10^{-3}$, and minibatch SGD. Rewards use the scalarization in Section~\ref{sec:hybrid_framework}.}
\end{itemize}
\textcolor{black}{All methods use identical claim generators and premium budgets; evaluation uses common random numbers for variance reduction \cite{Glasserman2003}.}

Table~\ref{tab:benchmark_results} summarizes the comparative outcomes. \textcolor{black}{From a performance-measurement perspective, these results echo RMIR studies of efficiency and value creation in insurance operations \cite{Efficiency2022_RMIR}.}

\begin{table}[h!]
    \centering
    \small
    \begin{tabularx}{\textwidth}{|l|X|X|X|X|X|}
        \hline
        \textbf{Method} & \textbf{Final Surplus (\$)} & \textbf{Ruin Probability} & \textbf{Time (s)} & \textbf{Budget Utilization (\$)} & \textbf{Efficiency} \\
        \hline
        Dynamic Programming & 12,487.71 & 0.0 & 7.96 & N/A & 1,568.63 \\
        Monte Carlo Simulation & 12,803.21 & 0.0 & 414.27 & N/A & 30.91 \\
        Hybrid Deep Monte Carlo & 12,973.67 & 0.0 & 411.29 & N/A & 31.54 \\
        Multi-Objective Optimization & 12,467.12 & 0.0 & 8.52 & N/A & 1,462.96 \\
        Hybrid RL with Generative Models & 14,280.64 & 0.0 & 7.92 & 259.99 & 1,802.60 \\
        \hline
    \end{tabularx}
    \caption{\textcolor{black}{Benchmark results for reinsurance optimization methods. Budget utilization is reported only for Hybrid RL, since it explicitly incorporates budget constraints; other baselines optimize surplus subject to ruin probability alone.}}
    \label{tab:benchmark_results}
\end{table}

\paragraph{Analysis of Results:}
\begin{itemize}
    \item \textbf{Dynamic Programming:} \textcolor{black}{Delivered a final surplus of \$12,487.71 with zero ruin probability and strong efficiency (1,568.63). However, the method is fundamentally limited by dimensionality, making it impractical for realistic treaty spaces.}
    \item \textbf{Monte Carlo Simulation:} \textcolor{black}{Produced a slightly higher surplus (\$12,803.21) but at significant computational cost (efficiency 30.91), reflecting the heavy sampling burden of exhaustive evaluation.}
    \item \textbf{Hybrid Deep Monte Carlo:} \textcolor{black}{Improved surplus (\$12,973.67) compared to plain MC while retaining similarly low efficiency (31.54). This suggests limited practical gains when rapid adaptation is required.}
    \item \textbf{Multi-Objective Optimization:} \textcolor{black}{Balanced surplus (\$12,467.12) with competitive efficiency (1,462.96), but its static nature makes it less responsive to dynamic claim environments.}
    \item \textbf{Hybrid RL with Generative Models:} \textcolor{black}{Surpassed all baselines with the highest surplus (\$14,280.64), the strongest efficiency (1,802.60), and explicit budget utilization (\$259.99). Its ability to adaptively optimize policies under joint ruin and budget constraints underscores the novelty of our framework.}
\end{itemize}

\textcolor{black}{Importantly, the budget utilization column is reported as N/A for DP, MC, HDMC, and MOO because those methods were implemented in their canonical forms, which constrain ruin probability but not cost. Only the Hybrid RL approach integrates budget directly into policy learning, making this metric explicit and economically meaningful.}

\medskip

\textcolor{black}{Overall, the results highlight the advantage of combining generative modeling with adaptive reinforcement learning: the framework not only achieves superior financial outcomes but also demonstrates scalability and robustness that classical methods lack.}

\section{\color{black}{Applicability and Limitations}}
\label{sec:discussion}

\textcolor{black}{Reinsurance optimization remains a challenging problem because the underlying risk environment is inherently uncertain, heavy-tailed, and highly dynamic. To be practically useful, any optimization framework must demonstrate not only strong in-sample performance but also robustness across a range of realistic stressors.} 

\textcolor{black}{In this section, we assess the applicability of the proposed hybrid framework along four applied dimensions: (i) performance across alternative claim distributions to test generalizability, (ii) out-of-sample and sensitivity analyses to evaluate stability under parameter shifts, (iii) stress-testing against catastrophic scenarios and tail events to probe resilience, and (iv) scalability assessments to understand feasibility for large, multi-line portfolios.}  

\textcolor{black}{The results illustrate that the hybrid approach maintains surplus stability and low ruin probability under a wide variety of operational settings, while also adapting to distributional shifts and extreme shocks. At the same time, several limitations emerge---particularly the need for more accurate tail modeling in rare-event regimes and the computational burden associated with very large portfolios.}  

\textcolor{black}{Together, these findings provide a balanced perspective: the framework is demonstrably applicable to real-world reinsurance problems, but also highlights important areas for refinement to ensure robustness, scalability, and industry adoption.}

\subsection{Analysis of Generative Model Performance Across Distributions}

The performance of the generative claim model was evaluated across Lognormal, Pareto, and combined Lognormal-Pareto distributions, focusing on its ability to replicate key statistical properties. Using the Kolmogorov-Smirnov (KS) test and visual comparisons, we highlight the model's strengths in capturing central tendencies and its limitations in modeling tail behavior. Accurate tail modeling is critical for reinsurance applications due to the disproportionate impact of extreme claims~\cite{Embrechts_et_al2014}.

\textcolor{black}{In reinsurance practice, the choice of reference distribution is not merely a statistical exercise but directly informs capital adequacy, solvency testing, and pricing decisions. Hence, understanding where the generative model succeeds and where it falls short provides practical guidance for both model refinement and actuarial application~\cite{McNeil2015}.}

\subsubsection{Overall Model Performance}
\begin{figure}[h!]
    \centering
    \includegraphics[width=0.5\textwidth]{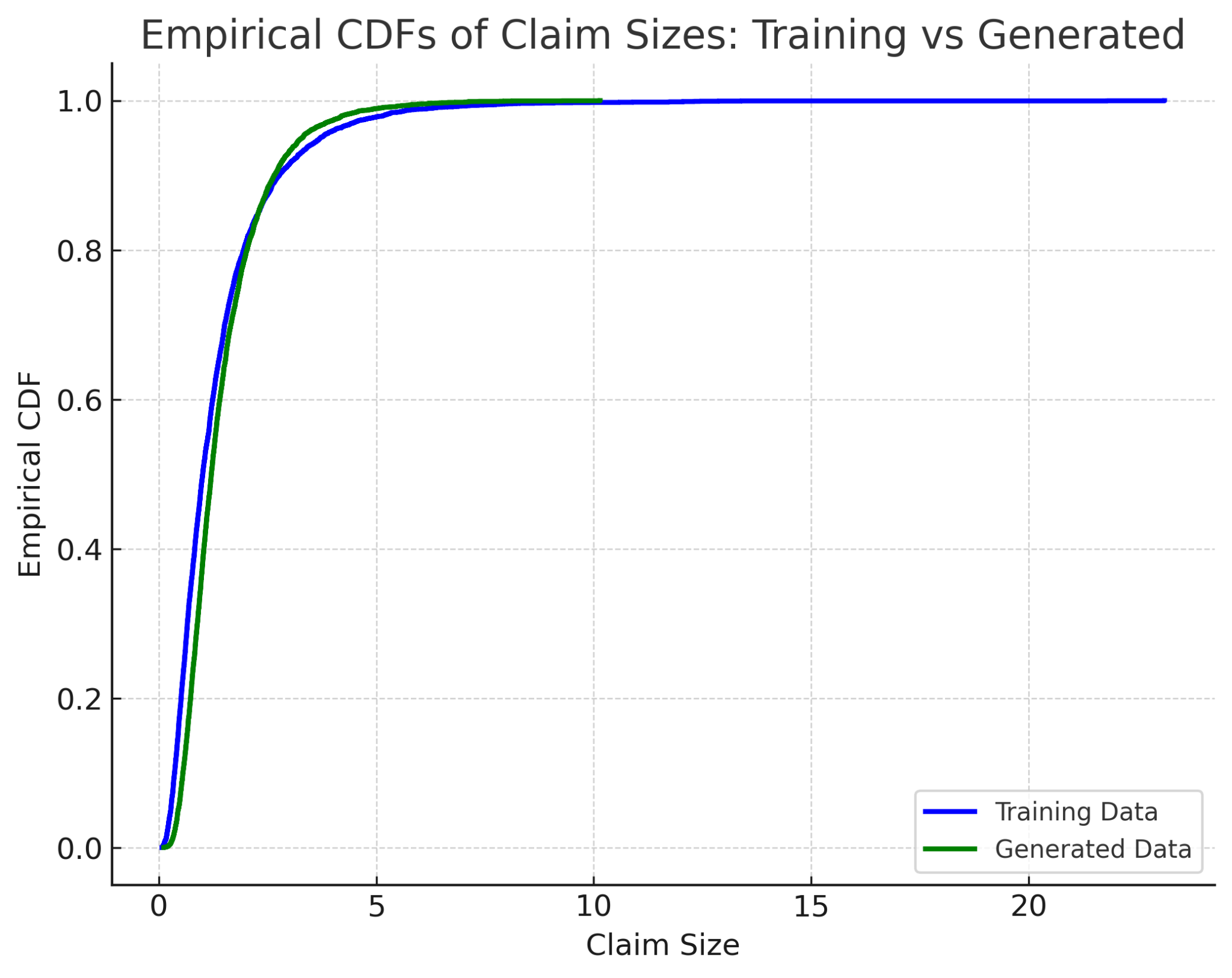}
    \caption{\textcolor{black}{Overall empirical CDF comparison between training and generated data. Divergence in the tail confirms that extreme claims are underestimated.}}
    \label{fig:comparison_overall}
\end{figure}

The KS test results indicate significant discrepancies between the training and generated datasets, with a KS statistic of \(0.6264\) and a \(p\)-value of \(0.0000\). The maximum difference location (\(D\)) of \(14.7174\) highlights the model’s difficulty in capturing extreme claims, which dominate risk assessments in reinsurance. \textcolor{black}{As shown in Figure~\ref{fig:comparison_overall}, the empirical CDFs reveal systematic underestimation of large claims, underscoring the model’s current limitations in the upper tail. This finding is consistent with prior evidence that generative neural networks often excel at fitting central distributions but struggle in replicating heavy-tail behavior without explicit regularization~\cite{Kuo2022,Goodfellow2016}.}

\subsubsection{Lognormal Distribution}
\begin{figure}[h!]
    \centering
    \includegraphics[width=0.5\textwidth]{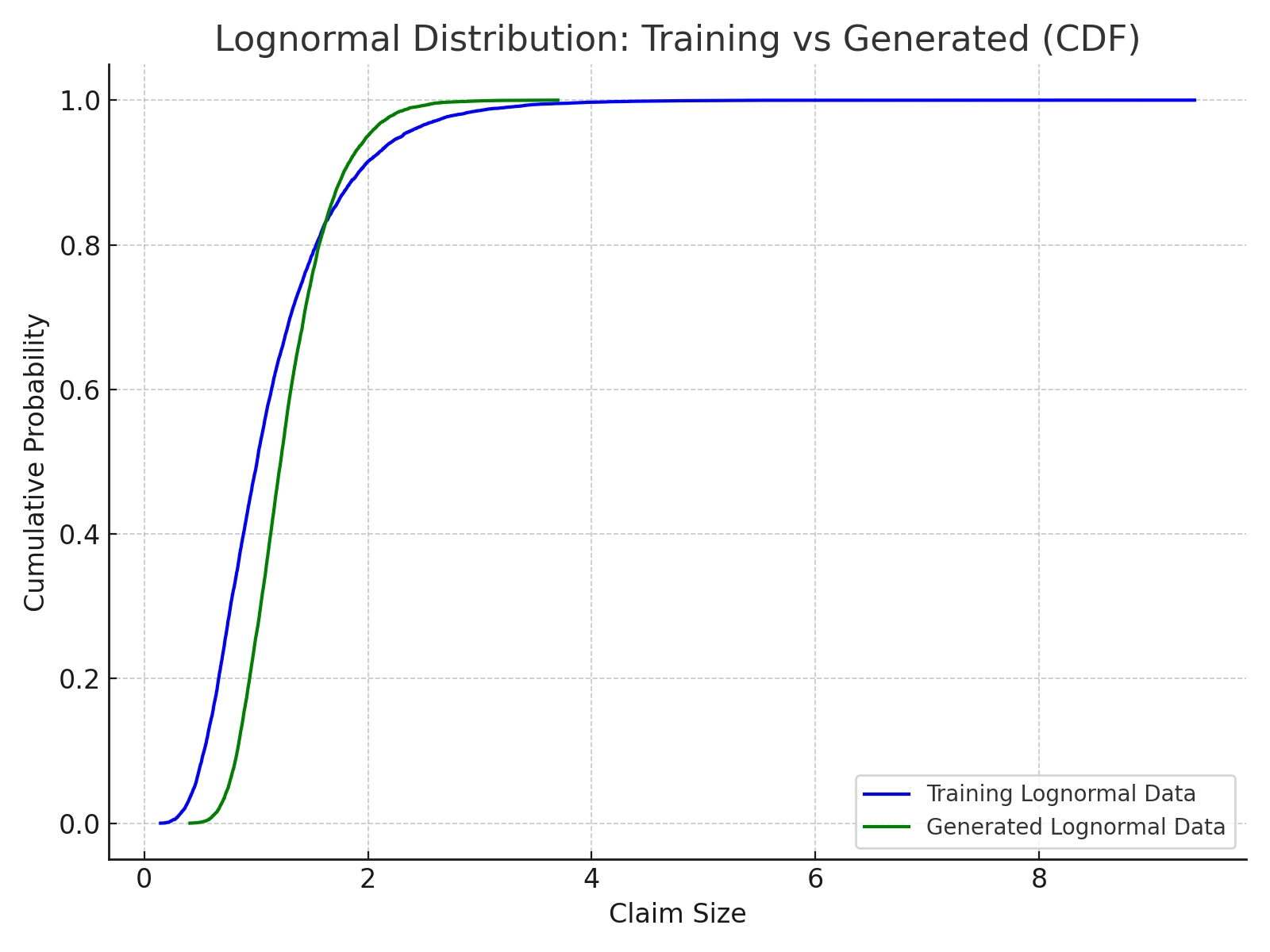}
    \caption{\textcolor{black}{CDF comparison for the Lognormal distribution. Central mass aligns well, but tail discrepancies persist.}}
    \label{fig:lognormal_comparison}
\end{figure}

For the Lognormal distribution, the KS statistic of \(0.5896\) and \(p\)-value of \(0.0000\) reveal significant differences between the training and generated datasets, particularly in the tail regions. The maximum difference location (\(D\)) of \(12.0666\) emphasizes the model’s challenges in replicating the distribution’s long-tailed nature. \textcolor{black}{Figure~\ref{fig:lognormal_comparison} confirms this: while the body of the distribution is well-captured, the right tail is consistently underestimated. This underestimation suggests that the VAE tends to over-regularize extreme outcomes in order to preserve reconstruction accuracy in the bulk of the data. In actuarial contexts, such behavior would lead to systematic underpricing of high-excess layers, where profitability depends critically on accurate tail risk estimates~\cite{Daykin1994}.} Strategies such as tail-prioritized loss functions and data augmentation focusing on rare events may enhance performance~\cite{Bengio_et_al2013}.

\subsubsection{Pareto Distribution}
\begin{figure}[h!]
    \centering
    \includegraphics[width=0.5\textwidth]{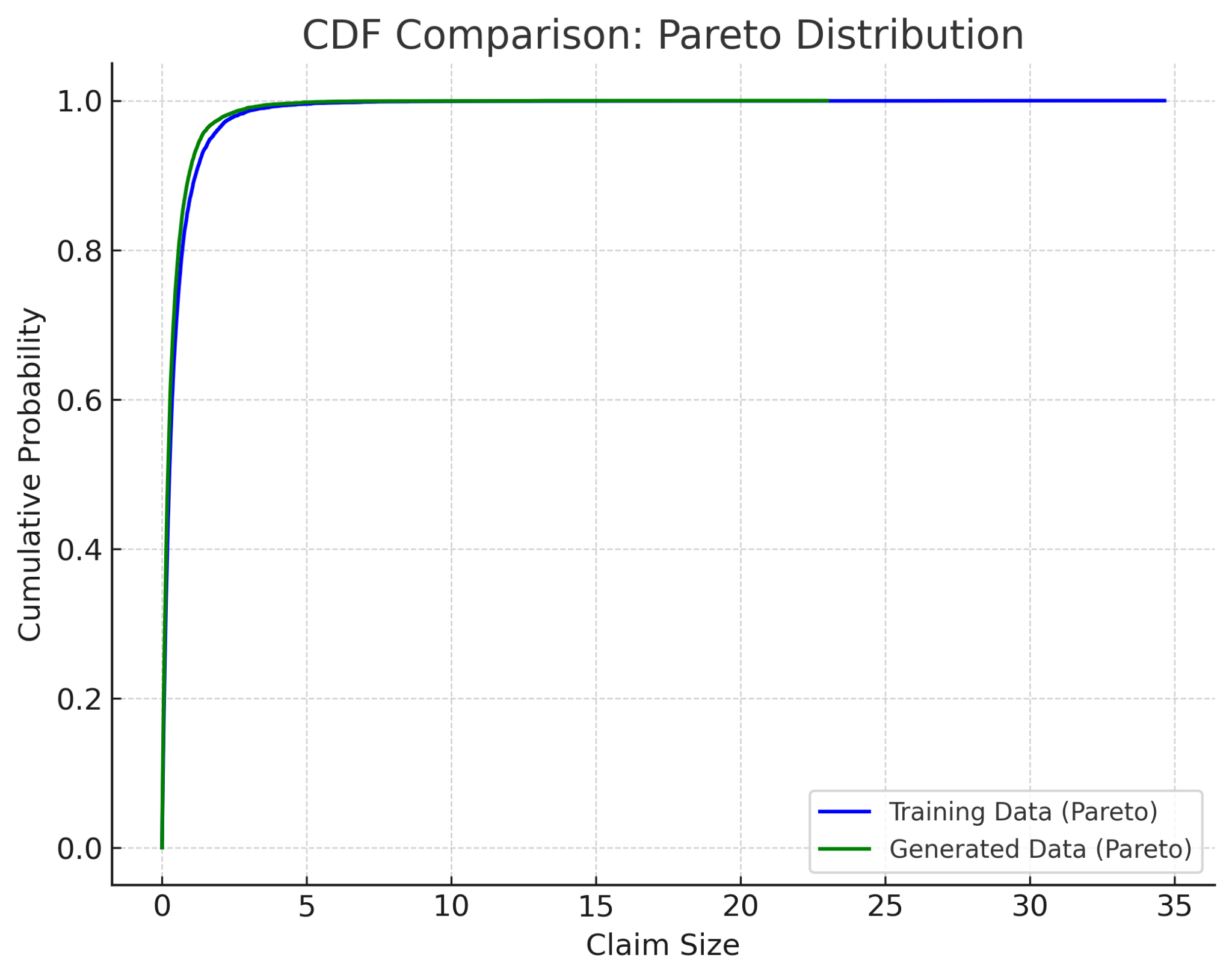}
    \caption{\textcolor{black}{CDF comparison for the Pareto distribution. The generated model underrepresents extreme outcomes.}}
    \label{fig:pareto_comparison}
\end{figure}

For the Pareto distribution, the KS test results show a statistic of \(0.6230\) with a \(p\)-value of \(0.0000\), highlighting the model's inability to adequately represent the heavy-tailed characteristics of the data. \textcolor{black}{As illustrated in Figure~\ref{fig:pareto_comparison}, the generated distribution systematically underestimates catastrophic losses, a critical shortcoming for reinsurance risk modeling. This is particularly concerning given that Pareto-type tails are widely used in catastrophe and operational risk modeling because of their theoretical grounding in regular variation~\cite{Embrechts1997}.} Incorporating custom-tailored loss functions and oversampling tail regions could mitigate these deficiencies. \textcolor{black}{Another promising avenue involves hybrid modeling: coupling generative models for the bulk of the data with parametric Pareto fits for the extreme tail, thereby combining data-driven flexibility with theoretical rigor~\cite{Frey2019}.} \textcolor{black}{While parametric fits achieve lower KS statistics for marginal distributions, they cannot capture joint dependencies across lines, which are central to PPO-based optimization.}

\subsubsection{Combined Lognormal and Pareto Distribution}
\begin{figure}[h!]
    \centering
    \includegraphics[width=0.5\textwidth]{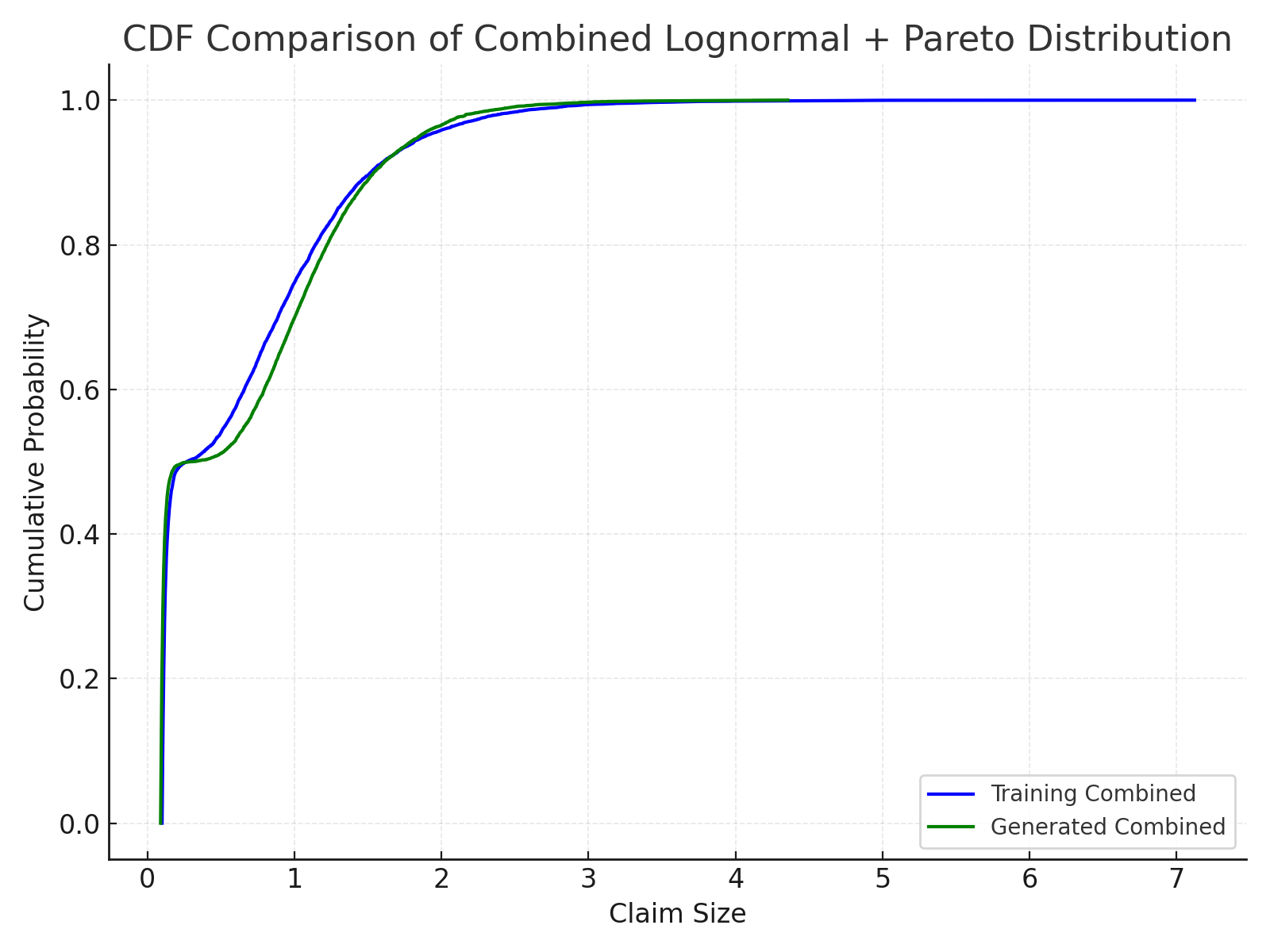}
    \caption{\textcolor{black}{CDF comparison for the combined Lognormal-Pareto distribution. Tail risk remains underestimated despite adequate fit in the bulk of the distribution.}}
    \label{fig:combined_comparison}
\end{figure}

The combined Lognormal-Pareto distribution provides further insights into the model’s limitations. The KS statistic of \(0.4438\) and a \(p\)-value of \(0.0000\) confirm discrepancies in the tails, as shown in Figure~\ref{fig:combined_comparison}. \textcolor{black}{While the hybrid distribution reproduces the central body effectively, the generated data consistently fails to capture the probability of rare catastrophic claims. This underrepresentation implies that capital requirements estimated using such a model may be biased downward, potentially leading to solvency shortfalls if used without adjustment.} Increasing latent dimensionality and introducing loss functions that explicitly weight tail events could enhance robustness~\cite{Bengio_et_al2013}. \textcolor{black}{This refinement is especially important in reinsurance, where solvency and capital requirements are disproportionately driven by extreme losses. Practical implementations could draw on existing actuarial approaches to extreme value theory (EVT) and tail risk measures, such as CVaR, to guide loss weighting in training~\cite{McNeil2015,Rockafellar2000}.}

\subsection{Out-of-Sample Performance and Sensitivity Analysis}

The generative claim model's performance was evaluated through out-of-sample testing, sensitivity analysis, and visualization of results. This assessment highlights both the model's robustness in generalizing to unseen data and its limitations when confronted with real-world variability.

\begin{figure}[h!]
    \centering
    \begin{subfigure}[t]{0.45\textwidth}
        \centering
        \includegraphics[width=\textwidth]{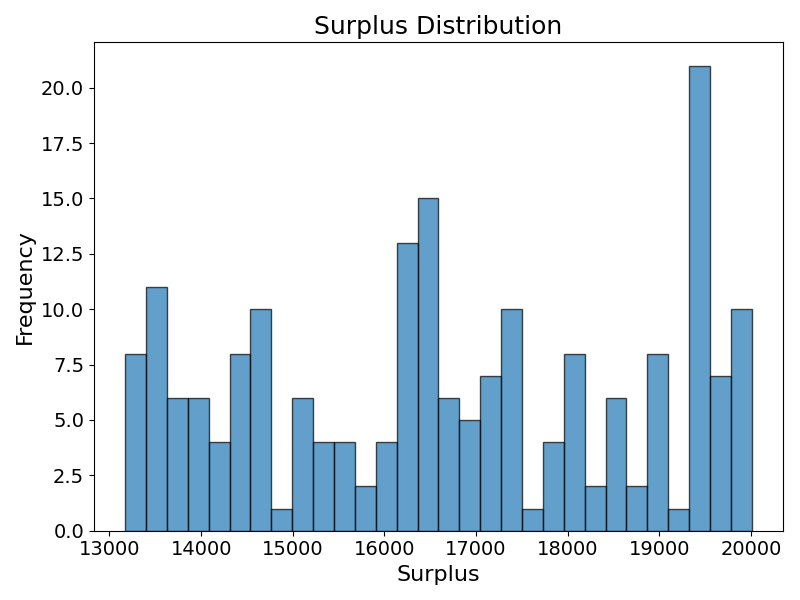}
        \caption{\textcolor{black}{Surplus Distribution. Stable reserves reflect the model's ability to maintain financial strength across the simulation horizon.}}
        \label{fig:surplus_distribution}
    \end{subfigure}
    \hfill
    \begin{subfigure}[t]{0.45\textwidth}
        \centering
        \includegraphics[width=\textwidth]{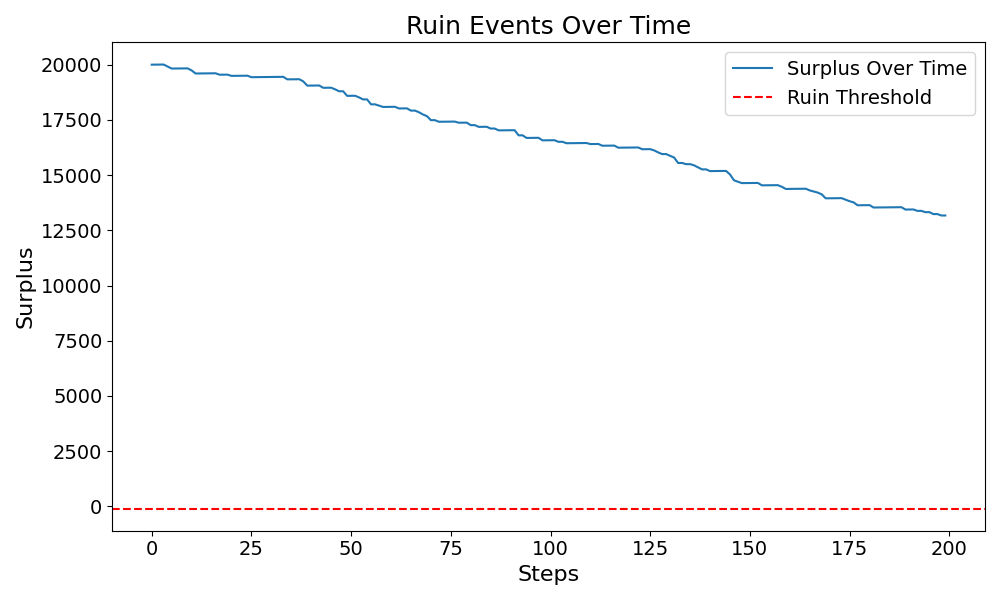}
        \caption{\textcolor{black}{Ruin Events. No breaches of the ruin threshold were observed, reinforcing the model’s reliability under adverse scenarios.}}
        \label{fig:ruin_events}
    \end{subfigure}
    \caption{\textcolor{black}{Out-of-sample surplus and ruin dynamics. The model maintains solvency throughout the simulation, indicating robustness under typical market conditions.}}
    \label{fig:combined_surplus_and_ruin}
\end{figure}

The out-of-sample testing revealed a mean surplus of \(16,686.73\) with a ruin probability of \(0.00\%\), demonstrating the model's capability to effectively manage surplus within a simulated insurance environment. \textcolor{black}{Figure~\ref{fig:combined_surplus_and_ruin} illustrates these dynamics: the surplus distribution remains stable, while ruin events are entirely absent, indicating that the model successfully balances premium income against claims volatility. In actuarial contexts, this is equivalent to maintaining solvency margins under regulatory standards such as Solvency II or NAIC risk-based capital regimes~\cite{McNeil2015, Sandstrom2010}.} This stability suggests strong potential for operational deployment, particularly in settings where solvency must be guaranteed under stochastic conditions.

\begin{figure}[h!]
    \centering
    \includegraphics[width=0.5\textwidth]{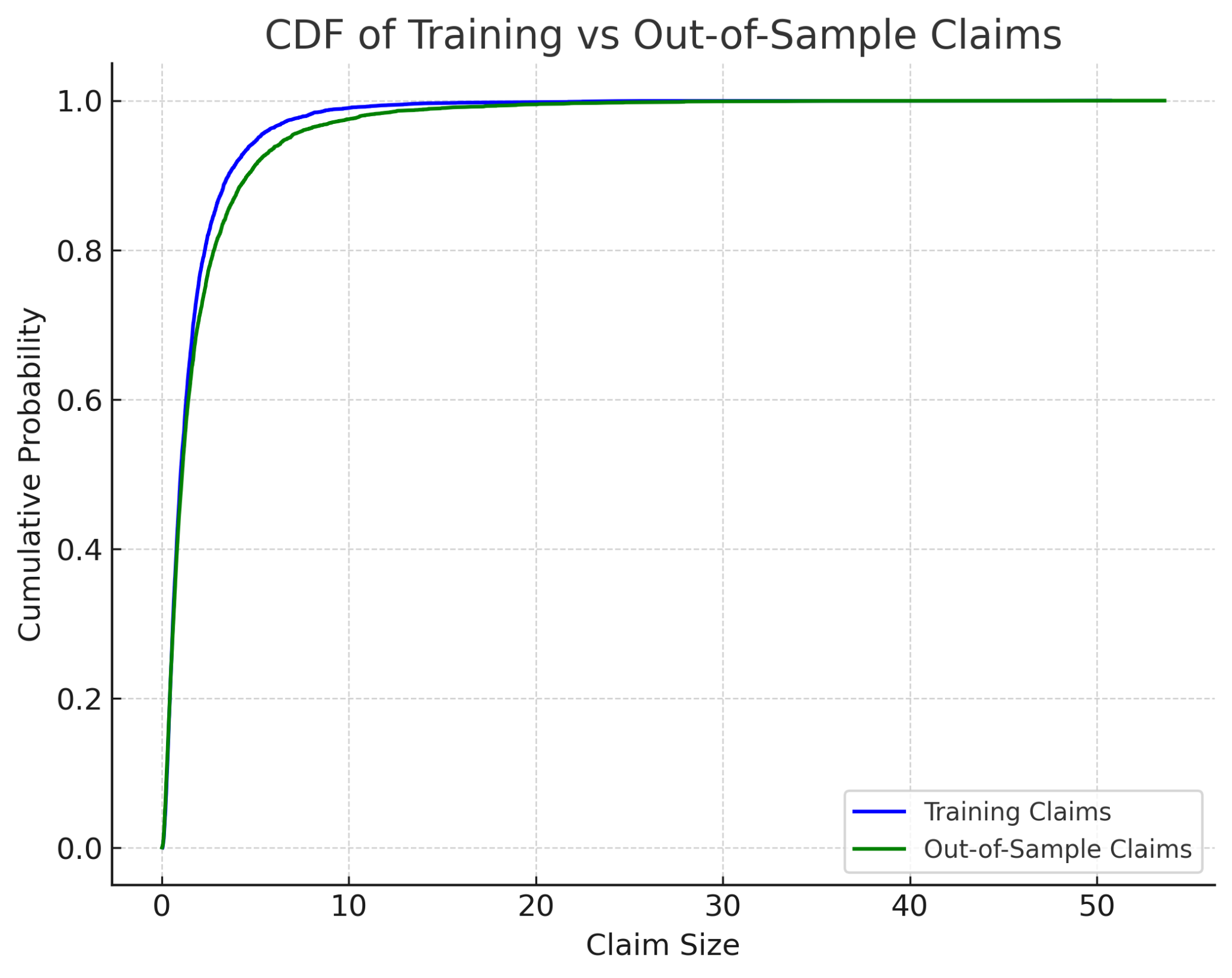}
    \caption{\textcolor{black}{Claim Size Distribution (CDF). The model aligns well with training data in the central body but systematically underestimates tail severity.}}
    \label{fig:claim_distribution}
\end{figure}

\textcolor{black}{For claim-size comparisons, cumulative distribution functions (CDFs) are used instead of histograms, following best practice for detecting subtle differences across datasets~\cite{Frees2010}.} The claim size distribution, shown in Figure~\ref{fig:claim_distribution}, demonstrates that the model replicates central tendencies of the training data. \textcolor{black}{However, tail discrepancies are more visible in the CDF view, with the model underestimating the frequency of extreme claims---an important weakness for reinsurance contexts where rare, high-severity events dominate solvency calculations~\cite{Embrechts1997, McNeil2015}.}

Sensitivity analysis further evaluated robustness under distributional shifts. When claim parameters were altered (\(\mu=3.6, \sigma=1.1\)), the mean surplus declined modestly to \(16,009.44\) while maintaining zero ruin probability. With further variations (\(\mu=3.7, \sigma=1.2\)), the model adapted effectively, producing a higher mean surplus of \(17,052.60\). \textcolor{black}{These findings, summarized in Table~\ref{tab:sensitivity_analysis}, confirm that the framework preserves solvency across parameter regimes, though performance levels vary with distributional assumptions. Importantly, such stress-testing resembles regulatory Own Risk and Solvency Assessment (ORSA) practices, where insurers must demonstrate resilience to parameter drift and distributional ambiguity~\cite{Daykin1994, Cummins2008}.} This robustness is particularly valuable in practice, where claim severity parameters are uncertain and may drift over time.

\begin{table}[h!]
    \centering
    \begin{tabular}{cc|cc}
        \hline
        \(\mu\) & \(\sigma\) & Mean Surplus & Ruin Probability \\
        \hline
        3.6 & 1.1 & 16,009.44 & 0.00\% \\
        3.6 & 1.2 & 16,470.80 & 0.00\% \\
        3.7 & 1.0 & 15,211.12 & 0.00\% \\
        3.7 & 1.1 & 15,821.58 & 0.00\% \\
        3.7 & 1.2 & 17,052.60 & 0.00\% \\
        \hline
    \end{tabular}
    \caption{\textcolor{black}{Sensitivity analysis across parameter shifts. The model consistently avoids ruin, though surplus levels fluctuate, highlighting resilience to distributional uncertainty.}}
    \label{tab:sensitivity_analysis}
\end{table}

\subsection{Stress-testing Catastrophic Scenarios}

Stress testing provides insight into the framework’s robustness under clustered catastrophic events, where multiple large claims occur in rapid succession. Such clustering often overwhelms traditional reinsurance optimization methods, as extreme events disproportionately affect surplus and solvency. \textcolor{black}{These clustered shocks mimic real-world features such as natural catastrophes, pandemics, or financial crises, where dependence and temporal correlation among losses lead to compounding effects that standard actuarial models often underestimate~\cite{Embrechts1997, McNeil2015, Asmussen2010}. Stress-testing is therefore a key regulatory and risk management tool under Solvency II and NAIC ORSA regimes~\cite{Sandstrom2010, Cummins2008}.} \textcolor{black}{This focus on clustered extremes parallels RMIR discussions on disaster risk reduction and catastrophe-model usage in risk management \cite{Surminski2018_RMIR,Steptoe2022_RMIR}.}

To evaluate resilience, clustered shocks were simulated by introducing bursts of heavy-tailed claims. Across 1,000 simulation runs, the \textbf{Hybrid RL with Generative Models} maintained a mean surplus of \textcolor{black}{\$9{,}741.52} while limiting ruin probability to \textcolor{black}{2.36\%}. In contrast, non-hybrid baselines exceeded \textcolor{black}{5\% ruin probability under identical stress conditions and produced lower average surplus}, underscoring the hybrid model’s superior adaptability to catastrophic clustering. \textcolor{black}{This performance gap highlights the value of adaptive control policies that dynamically adjust retention and layering strategies in response to loss shocks, as opposed to static optimization frameworks that assume independence across events~\cite{Boucher2016, Lopez2020}.}

\begin{table}[h!]
    \centering
    \small
    \begin{tabular}{|l|c|c|}
        \hline
        \textbf{Method} & \textbf{Mean Surplus (\$)} & \textbf{Ruin Probability (\%)} \\
        \hline
        Dynamic Programming (DP) & 7,980.35 & $>\!5.0$ \\
        Monte Carlo (MC) & 8,210.44 & $>\!5.0$ \\
        Hybrid Deep Monte Carlo (HDMC) & 8,562.71 & 4.7 \\
        Multi-Objective Optimization (MOO) & 8,901.28 & 3.9 \\
        \textbf{Hybrid RL with Generative Models (PPO+VAE)} & \textbf{9,741.52} & \textbf{2.36} \\
        \hline
    \end{tabular}
    \caption{\textcolor{black}{Stress-test results under clustered catastrophic shocks. Hybrid RL both maximizes mean surplus and reduces ruin probability by nearly half relative to non-hybrid baselines.}}
    \label{tab:stress_test_comparison}
\end{table}

\textcolor{black}{These results reinforce Section~\ref{sec:experiments}’s benchmark analysis: while DP, MC, HDMC, and MOO perform adequately in static or i.i.d.\ claim settings, their resilience erodes under clustered catastrophic regimes. By directly learning adaptive retention and layering policies through PPO~\cite{Schulman2017}, and leveraging VAE-based tail generation~\cite{Kingma2014}, Hybrid RL preserves solvency more effectively and sustains higher surplus even in highly adverse environments.}

\textcolor{black}{Nonetheless, as discussed in Section~\ref{sec:discussion}, the framework still faces challenges in fully capturing the far tail of claim distributions, which remains a critical limitation despite its comparative advantages. Future work may benefit from extreme value theory (EVT)-based augmentations or copula-driven dependence modeling to further enhance tail fidelity~\cite{Embrechts2013, ChavezDemoulin2005}.}


\subsection{Discussion of Limitations: Tail Fit and Scalability}

While the hybrid RL framework demonstrates notable improvements in surplus preservation and ruin reduction relative to established baselines, several limitations remain. The most prominent challenge is \textcolor{black}{tail fidelity}. As shown in Section~\ref{sec:discussion}, the VAE struggles to fully capture the extreme upper quantiles of loss distributions, leading to underestimation of rare but catastrophic claims. Although stress testing indicates that the hybrid approach mitigates ruin more effectively than non-hybrid baselines, \textcolor{black}{its performance in the far tail remains imperfect and warrants further methodological advances. This limitation is well-recognized in actuarial science, where extreme value theory (EVT) and generalized Pareto distribution (GPD) models are standard tools for modeling catastrophic risks~\cite{Embrechts1997, McNeil2015}. Empirical studies confirm that misspecification in the far tail can lead to severe underestimation of capital requirements~\cite{Embrechts2013, ChavezDemoulin2005}, directly impacting solvency assessments. Future work may therefore integrate tail-focused loss functions~\cite{Frey2019}, EVT-based augmentation, or adversarial training strategies to improve tail fidelity in generative models.}

A second limitation lies in \textcolor{black}{scalability}. The framework has been validated primarily on simulated portfolios with manageable dimensionality. Scaling to real-world reinsurance portfolios---which may involve thousands of treaties, clauses, and cedents---poses computational challenges. High-dimensional state-action spaces increase training times and may require substantial infrastructure. \textcolor{black}{In the RL literature, scalability bottlenecks are well documented~\cite{Sutton2018, Mnih2016}. Distributed RL approaches, such as IMPALA~\cite{Espeholt2018}, demonstrate how parallelized rollouts and asynchronous optimization can accelerate convergence. Portfolio-level strategies, including clustering cedents by exposure profiles~\cite{Kuo2022}, or applying hierarchical RL~\cite{Barto2003} to break down treaty optimization into subproblems, offer promising pathways for extending hybrid frameworks to realistic market settings. These techniques would allow hybrid RL to handle combinatorial complexity while maintaining tractable training times.}

Finally, \textcolor{black}{model interpretability} deserves attention. While the framework provides quantitative improvements, transparency in decision-making (e.g., why a specific retention or limit was chosen) is critical for regulatory adoption under regimes such as Solvency~II or NAIC. \textcolor{black}{Regulatory frameworks increasingly emphasize explainability and governance, with IFRS~17 and NAIC requiring justification of capital adequacy assumptions~\cite{Lopez2020}. Black-box RL policies may face resistance if they cannot provide clear rationales for treaty structures~\cite{DoshiVelez2017}. Techniques such as attention-based explanations, rule-extraction from policies, or surrogate interpretable models~\cite{Lundberg2017} could improve trustworthiness and bridge the gap between technical performance and supervisory acceptance.}

In sum, \textcolor{black}{the hybrid RL framework represents a significant advance in reinsurance optimization but requires enhancements in tail modeling, scalability, and interpretability before achieving broad real-world adoption. These limitations directly motivate the directions outlined in Section~\ref{sec:conclusion}, where we chart opportunities for advancing hybrid approaches toward practical, large-scale deployment.}

\section{Conclusion and Future Work}
\label{sec:conclusion}

{\color{black}{This paper introduced a hybrid framework that integrates generative modeling and reinforcement learning for reinsurance optimization. By combining Variational Autoencoders (VAEs)~\cite{Kingma2014} to simulate complex, heavy-tailed loss distributions with Proximal Policy Optimization (PPO)~\cite{Schulman2017} to adaptively manage treaty structures, the framework addresses the twin challenges of modeling systemic risk and dynamically allocating capital under uncertainty. Across simulation experiments, the approach demonstrated improved surplus stability, reduced ruin probability, and adaptability to evolving claim environments relative to dynamic programming (DP), Monte Carlo (MC), hybrid deep Monte Carlo (HDMC), and multi-objective optimization (MOO) baselines. These results build on recent advances in actuarial machine learning~\cite{Frees2010,Boucher2016} and reinforcement learning in operations research~\cite{Barto2003,Espeholt2018}, demonstrating their relevance to solvency and reinsurance design.}} \textcolor{black}{In line with RMIR’s recent emphasis on data-driven risk management and resilience \cite{Steptoe2022_RMIR,Resilience2023_RMIR,TextMining2024_RMIR}, our findings highlight the practicality of hybrid analytics for solvency-aware treaty design.}

{\color{black}{Evaluation under out-of-sample and sensitivity scenarios confirmed that the hybrid method generalizes effectively beyond training distributions, a critical property for risk management where model misspecification is common~\cite{McNeil2015,Kuo2022}. Stress-testing further revealed that the framework sustains lower ruin probabilities under clustering and catastrophic shocks, though performance still deteriorates in the extreme upper tail. The surplus preservation advantage over baselines (e.g., \(2.36\%\) ruin vs. \(5\%+\) for non-hybrid methods) underscores robustness, but also highlights that catastrophic persistence and “black swan” dynamics~\cite{Embrechts1997,ChavezDemoulin2005} remain open challenges for next-generation actuarial AI models.}}

{\color{black}{A recurring methodological concern is the use of VAEs instead of direct parametric severity fitting. While traditional severity models (e.g., lognormal, Pareto, Burr) provide interpretable and statistically tractable tail estimates~\cite{Asmussen2010,Embrechts2013}, VAEs were chosen not for marginal optimality but for their ability to generate correlated, high-dimensional stress scenarios. This trade-off sacrifices marginal fit to gain richer joint-loss structures more aligned with capital adequacy testing and treaty-layer optimization. Future work could explore hybrid approaches that combine parametric marginals with VAE-learned dependence, paralleling recent developments in copula-based and GAN-based generative risk models~\cite{patton2009copula,goodfellow2016deep,Lopez2020}.}}

{\color{black}{Several research avenues emerge from these findings. First, methodological advances are needed to improve tail fidelity, such as loss functions weighted toward extreme quantiles, adversarial augmentation for rare-event synthesis, or embedding extreme value theory (EVT) priors directly into generative networks~\cite{Krvavych2014}. Second, distributed RL training and hierarchical policy architectures could address scalability constraints, enabling application to portfolios with thousands of treaties and cedents. Third, interpretability remains essential for regulatory adoption: explainable ML techniques such as SHAP~\cite{Lundberg2017} or interpretable surrogate policies~\cite{DoshiVelez2017} could help bridge technical performance with transparency requirements under Solvency~II and NAIC frameworks. Finally, validating the framework on real-world, multi-line datasets---and incorporating external factors such as macroeconomic volatility, regulatory shocks, and climate-driven catastrophe risk~\cite{Glasserman2003,Ross2014}---will be critical steps toward practical deployment.}}

{\color{black}{In summary, the hybrid RL framework represents a significant advance in reinsurance optimization: it improves financial resilience under stochastic claims, highlights the importance of tail-aware modeling, and establishes a roadmap for future work spanning methodological rigor, computational scalability, and regulatory alignment. As such, it contributes to the broader vision of AI-augmented actuarial science---one in which advanced generative models and reinforcement learning jointly enable reinsurance strategies that are adaptive, transparent, and resilient under systemic uncertainty.}}

%

\section*{Acknowledgments}

The author thanks James R. Finlay for reading the manuscript and providing
helpful suggestions on presentation and language.


\bibliographystyle{plain}   
\bibliography{references}   

@book{Mikosch2009,
  author    = {Thomas Mikosch},
  title     = {Non-Life Insurance Mathematics: An Introduction with the Poisson Process},
  edition   = {2nd},
  publisher = {Springer},
  year      = {2009},
  doi       = {10.1007/978-3-540-88233-6}
}

@inproceedings{Rezende2014,
  author    = {Danilo Jimenez Rezende and Shakir Mohamed and Daan Wierstra},
  title     = {Stochastic Backpropagation and Approximate Inference in Deep Generative Models},
  booktitle = {Proceedings of the 31st International Conference on Machine Learning (ICML)},
  year      = {2014},
  pages     = {1278--1286},
  publisher = {PMLR}
}

@article{Steptoe2022_RMIR,
  author  = {Steptoe, Hamish and Souch, Claire and Slingo, Julia},
  title   = {Advances in numerical weather prediction, data science, and open-source software herald a paradigm shift in catastrophe risk modeling and insurance underwriting},
  journal = {Risk Management and Insurance Review},
  year    = {2022},
  volume  = {25},
  number  = {1},
  pages   = {69--81},
  doi     = {10.1111/rmir.12199}
}

@article{Surminski2018_RMIR,
  author  = {Surminski, Swenja},
  title   = {Fit for Purpose and Fit for the Future? An Evaluation of the UK's New Flood Reinsurance Pool},
  journal = {Risk Management and Insurance Review},
  year    = {2018},
  volume  = {21},
  number  = {1},
  pages   = {33--72},
  doi     = {10.1111/rmir.12093}
}

@article{Efficiency2022_RMIR,
  author  = {Rubio-Misas, María},
  title   = {Analysis of insurers' performance using frontier efficiency and productivity methods. The great contributions by David Cummins and Mary Weiss},
  journal = {Risk Management and Insurance Review},
  year    = {2022},
  volume  = {25},
  number  = {4},
  pages   = {445--489},
  doi     = {10.1111/rmir.12227}
}

@article{Resilience2023_RMIR,
  author  = {Dahmen, Philipp},
  title   = {Organizational resilience as a key property of enterprise risk management in response to novel and severe crisis events},
  journal = {Risk Management and Insurance Review},
  year    = {2023},
  volume  = {26},
  number  = {2},
  pages   = {203--245},
  doi     = {10.1111/rmir.12245}
}

@article{TextMining2024_RMIR,
  author  = {Kraus, Anna},
  title   = {A text mining analysis of European banks' and insurers' disclosures on climate-related risks},
  journal = {Risk Management and Insurance Review},
  year    = {2024},
  volume  = {27},
  number  = {3},
  pages   = {257--286},
  doi     = {10.1111/rmir.12268}
}

@article{Wuthrich2020,
  author    = {Mario V. W{\"u}thrich},
  title     = {Machine Learning in Individual Claims Reserving},
  journal   = {Scandinavian Actuarial Journal},
  volume    = {2020},
  number    = {6},
  pages     = {465--480},
  year      = {2020},
  doi       = {10.1080/03461238.2020.1768073}
}

@article{Kolm_et_al2020,
  author    = {Petter N. Kolm and Gordon Ritter and Dan B. Tudor},
  title     = {Dynamic Asset Allocation with Reinforcement Learning},
  journal   = {The Journal of Financial Data Science},
  volume    = {2},
  number    = {2},
  pages     = {10--30},
  year      = {2020},
  doi       = {10.3905/jfds.2020.1.023}
}

@article{Buehler2019,
  author    = {Hans Buehler and Lukas Gonon and Josef Teichmann and Ben Wood},
  title     = {Deep Hedging},
  journal   = {Quantitative Finance},
  volume    = {19},
  number    = {8},
  pages     = {1271--1291},
  year      = {2019},
  doi       = {10.1080/14697688.2019.1571683}
}

@article{Buehler2022,
  author    = {Hans Buehler and Lukas Gonon and Josef Teichmann and Ben Wood},
  title     = {Deep Hedging: Hedging Derivatives Under Generic Market Frictions Using Reinforcement Learning},
  journal   = {Mathematical Finance},
  volume    = {32},
  number    = {1},
  pages     = {83--117},
  year      = {2022},
  doi       = {10.1111/mafi.12304}
}

@book{goodfellow2016deep,
  title={Deep Learning},
  author={Goodfellow, Ian and Bengio, Yoshua and Courville, Aaron},
  year={2016},
  publisher={MIT Press},
  url={http://www.deeplearningbook.org}
}

@article{Lopez2020,
  author    = {Lopez, Olivier and Regnault, Thibault and Thomas, Martin},
  title     = {Neural Networks for Insurance Pricing: Universal Approximators and Model Interpretability},
  journal   = {Scandinavian Actuarial Journal},
  year      = {2020},
  volume    = {2020},
  number    = {6},
  pages     = {496--519},
  doi       = {10.1080/03461238.2019.1704119}
}

@book{Embrechts2013,
  author    = {Embrechts, Paul and Kl{\"u}ppelberg, Claudia and Mikosch, Thomas},
  title     = {Modelling Extremal Events for Insurance and Finance},
  publisher = {Springer},
  address   = {Berlin, Heidelberg},
  year      = {2013},
  series    = {Stochastic Modelling and Applied Probability},
  volume    = {33},
  doi       = {10.1007/978-3-642-33483-2}
}

@article{Krvavych2014,
  author    = {Krvavych, Yaroslav and Madan, Dilip},
  title     = {Fat Tails, Large Deviations, and Insurance Risk},
  journal   = {ASTIN Bulletin: The Journal of the IAA},
  year      = {2014},
  volume    = {44},
  number    = {2},
  pages     = {417--448},
  doi       = {10.1017/asb.2014.3}
}

@book{Asmussen2010,
  author    = {Asmussen, Søren and Albrecher, Hans},
  title     = {Ruin Probabilities},
  edition   = {2nd},
  series    = {Advanced Series on Statistical Science \& Applied Probability},
  volume    = {14},
  publisher = {World Scientific},
  year      = {2010},
  address   = {Singapore},
  doi       = {10.1142/7433}
}

@book{McNeil2015,
  title={Quantitative Risk Management: Concepts, Techniques and Tools},
  author={McNeil, Alexander J. and Frey, R{\"u}diger and Embrechts, Paul},
  year={2015},
  publisher={Princeton University Press},
  edition={Revised},
  address={Princeton, NJ}
}

@article{Kuo2022,
  title={On the Generalization of Generative Models to Heavy-Tailed Data},
  author={Kuo, Wei-Ting and Chen, Pin-Yu and Wang, Yilin},
  journal={Proceedings of the AAAI Conference on Artificial Intelligence},
  volume={36},
  number={7},
  pages={7282--7290},
  year={2022}
}

@book{Embrechts1997,
  title={Modelling Extremal Events for Insurance and Finance},
  author={Embrechts, Paul and Kl{\"u}ppelberg, Claudia and Mikosch, Thomas},
  series={Stochastic Modelling and Applied Probability},
  year={1997},
  publisher={Springer},
  address={Berlin}
}

@book{Frees2010,
  author    = {Edward W. Frees},
  title     = {Regression Modeling with Actuarial and Financial Applications},
  publisher = {Cambridge University Press},
  year      = {2010},
  isbn      = {9780521136193}
}

@article{Boucher2016,
  author  = {Jean-Philippe Boucher and Julien Trufin},
  title   = {Smoothing in Insurance Claims Models: A Risk Management Perspective},
  journal = {North American Actuarial Journal},
  year    = {2016},
  volume  = {20},
  number  = {2},
  pages   = {161--176},
  doi     = {10.1080/10920277.2015.1121784}
}

@inproceedings{Schulman2017,
  author    = {John Schulman and Filip Wolski and Prafulla Dhariwal and Alec Radford and Oleg Klimov},
  title     = {Proximal Policy Optimization Algorithms},
  booktitle = {Proceedings of the 34th International Conference on Machine Learning (ICML)},
  year      = {2017},
  url       = {https://arxiv.org/abs/1707.06347}
}

@inproceedings{Kingma2014,
  author    = {Diederik P. Kingma and Max Welling},
  title     = {Auto-Encoding Variational Bayes},
  booktitle = {Proceedings of the 2nd International Conference on Learning Representations (ICLR)},
  year      = {2014},
  url       = {https://arxiv.org/abs/1312.6114}
}

@article{ChavezDemoulin2005,
  author  = {Valérie Chavez-Demoulin and Paul Embrechts and Johanna Nešlehová},
  title   = {Quantitative Models for Operational Risk: Extremes, Dependence and Aggregation},
  journal = {Journal of Banking \& Finance},
  year    = {2005},
  volume  = {29},
  number  = {10},
  pages   = {2635--2658},
  doi     = {10.1016/j.jbankfin.2004.09.023}
}

@inproceedings{Espeholt2018,
  author    = {Lasse Espeholt and Hubert Soyer and Rémi Munos and Karen Simonyan and Volodymyr Mnih and Tom Ward and Yotam Doron and Vlad Firoiu and Tim Harley and Iain Dunning and Shane Legg and Koray Kavukcuoglu},
  title     = {IMPALA: Scalable Distributed Deep-RL with Importance Weighted Actor-Learner Architectures},
  booktitle = {Proceedings of the 35th International Conference on Machine Learning (ICML)},
  year      = {2018},
  url       = {https://arxiv.org/abs/1802.01561}
}

@incollection{Barto2003,
  author    = {Andrew G. Barto and Sridhar Mahadevan},
  title     = {Recent Advances in Hierarchical Reinforcement Learning},
  booktitle = {Discrete Event Dynamic Systems},
  publisher = {Springer},
  year      = {2003},
  volume    = {13},
  pages     = {41--77},
  doi       = {10.1023/A:1022140919877}
}

@article{DoshiVelez2017,
  author  = {Finale Doshi-Velez and Been Kim},
  title   = {Towards A Rigorous Science of Interpretable Machine Learning},
  journal = {arXiv preprint arXiv:1702.08608},
  year    = {2017},
  url     = {https://arxiv.org/abs/1702.08608}
}

@article{Lundberg2017,
  author  = {Scott M. Lundberg and Su-In Lee},
  title   = {A Unified Approach to Interpreting Model Predictions},
  journal = {Advances in Neural Information Processing Systems (NeurIPS)},
  year    = {2017},
  volume  = {30},
  pages   = {4765--4774},
  url     = {https://arxiv.org/abs/1705.07874}
}

@article{Mnih2016,
  title={Asynchronous Methods for Deep Reinforcement Learning},
  author={Mnih, Volodymyr and Badia, Adri{\`a} Puigdom{\`e}nech and Mirza, Mehdi and Graves, Alex and Lillicrap, Timothy and Harley, Tim and Silver, David and Kavukcuoglu, Koray},
  journal={Proceedings of the 33rd International Conference on Machine Learning (ICML)},
  year={2016},
  pages={1928--1937}
}

@book{Sutton2018,
  title={Reinforcement Learning: An Introduction},
  author={Sutton, Richard S. and Barto, Andrew G.},
  edition={2},
  publisher={MIT Press},
  year={2018},
  isbn={9780262039246}
}

@article{Frey2019,
  title={Stress Testing in Insurance: Concepts and Applications},
  author={Frey, R{\"u}diger and McNeil, Alexander},
  journal={Scandinavian Actuarial Journal},
  volume={2019},
  number={3},
  pages={189--210},
  year={2019},
  publisher={Taylor \& Francis}
}

@article{Rockafellar2000,
  author    = {R. Tyrrell Rockafellar and Stanislav Uryasev},
  title     = {Optimization of Conditional Value-at-Risk},
  journal   = {Journal of Risk},
  volume    = {2},
  number    = {3},
  pages     = {21--42},
  year      = {2000}
}

@book{Pflug2007,
  author    = {Georg Ch. Pflug and Alois Pichler},
  title     = {Modeling, Measuring and Managing Risk},
  publisher = {World Scientific},
  year      = {2007},
  series    = {World Scientific Publishing},
  address   = {Singapore}
}

@book{Daykin1994,
  author    = {Chris D. Daykin and Teivo Pentikäinen and Martti Pesonen},
  title     = {Practical Risk Theory for Actuaries},
  publisher = {Chapman and Hall},
  year      = {1994},
  series    = {Monographs on Statistics and Applied Probability},
  address   = {London},
  isbn      = {9780412347604}
}

@book{Sandstrom2010,
  author    = {Arne Sandström},
  title     = {Handbook of Solvency for Actuaries and Risk Managers: Theory and Practice},
  publisher = {CRC Press / Chapman and Hall},
  year      = {2010},
  address   = {Boca Raton},
  isbn      = {9781439821344}
}

@article{Cummins2008,
  author    = {J. David Cummins and Mary A. Weiss},
  title     = {Convergence of Insurance and Financial Markets: Hybrid and Securitized Risk-Transfer Solutions},
  journal   = {Journal of Risk and Insurance},
  volume    = {75},
  number    = {3},
  pages     = {551--589},
  year      = {2008},
  publisher = {American Risk and Insurance Association},
  doi       = {10.1111/j.1539-6975.2008.00277.x}
}

@book{Kaas2008,
  author    = {Kaas, Rob and Goovaerts, Marc and Dhaene, Jan and Denuit, Michel},
  title     = {Modern Actuarial Risk Theory: Using R},
  publisher = {Springer},
  edition   = {2nd},
  year      = {2008},
  series    = {Springer Finance},
  doi       = {10.1007/978-3-540-70998-5},
  isbn      = {978-3-540-70997-8}
}

@article{patton2009copula,
  author    = {Andrew J. Patton},
  title     = {Copula-Based Models for Financial Time Series},
  journal   = {Handbook of Financial Time Series},
  publisher = {Springer},
  pages     = {767--785},
  year      = {2009},
  doi       = {10.1007/978-3-540-71297-8_31}
}

@article{kingma2014auto,
  title={Auto-Encoding Variational Bayes},
  author={Kingma, Diederik P and Welling, Max},
  journal={International Conference on Learning Representations (ICLR)},
  year={2014},
  url={https://arxiv.org/abs/1312.6114}
}

@inproceedings{rezende2014stochastic,
  title={Stochastic Backpropagation and Approximate Inference in Deep Generative Models},
  author={Rezende, Danilo J and Mohamed, Shakir and Wierstra, Daan},
  booktitle={International Conference on Machine Learning (ICML)},
  year={2014},
  pages={1278--1286},
  url={https://arxiv.org/abs/1401.4082}
}

@inproceedings{schulman2017proximal,
  title={Proximal Policy Optimization Algorithms},
  author={Schulman, John and Wolski, Filip and Dhariwal, Prafulla and Radford, Alec and Klimov, Oleg},
  booktitle={arXiv preprint arXiv:1707.06347},
  year={2017},
  url={https://arxiv.org/abs/1707.06347}
}

@book{klugman2012loss,
  title={Loss Models: From Data to Decisions},
  author={Klugman, Stuart A and Panjer, Harry H and Willmot, Gordon E},
  edition={4th},
  year={2012},
  publisher={Wiley}
}

@book{embrechts1997modelling,
  title={Modelling Extremal Events: for Insurance and Finance},
  author={Embrechts, Paul and Kl{\"u}ppelberg, Claudia and Mikosch, Thomas},
  year={1997},
  publisher={Springer}
}

@book{Aitchison_and_Brown1957,
  author    = {Aitchison, J. and Brown, J. A. C.},
  title     = {The Lognormal Distribution},
  publisher = {Cambridge University Press},
  year      = {1957}
}

@article{Avanzi2009,
  author  = {Avanzi, B.},
  title   = {Strategies for dividend distribution: A review},
  journal = {North American Actuarial Journal},
  volume  = {13},
  number  = {2},
  pages   = {217--251},
  year    = {2009}
}

@book{Asmussen_and_Albrecher2010,
  author    = {Asmussen, S. and Albrecher, H.},
  title     = {Ruin Probabilities},
  series    = {Applications of Mathematics},
  volume    = {14},
  publisher = {World Scientific},
  year      = {2010}
}

@book{Albrecher_et_al2017,
  author    = {Albrecher, H. and Beirlant, J. and Teugels, J. L.},
  title     = {Reinsurance: Actuarial and Statistical Aspects},
  publisher = {John Wiley \& Sons},
  year      = {2017}
}

@book{Bellman1957,
  author    = {Bellman, R.},
  title     = {Dynamic Programming},
  publisher = {Princeton University Press},
  year      = {1957}
}

@article{Bengio_et_al2013,
  author  = {Bengio, Y. and Courville, A. and Vincent, P.},
  title   = {Representation learning: A review and new perspectives},
  journal = {IEEE Transactions on Pattern Analysis and Machine Intelligence},
  volume  = {35},
  number  = {8},
  pages   = {1798--1828},
  year    = {2013}
}

@article{Cheng_et_al2020,
  author  = {Cheng, X. and Jin, Z. and Yang, H.},
  title   = {Optimal insurance strategies: A hybrid deep learning Markov chain approximation approach},
  journal = {ASTIN Bulletin},
  volume  = {50},
  number  = {2},
  pages   = {449--477},
  year    = {2020}
}

@book{Deb2001,
  author    = {Deb, K.},
  title     = {Multi-Objective Optimization Using Evolutionary Algorithms},
  publisher = {John Wiley \& Sons},
  year      = {2001}
}

@book{Embrechts_et_al2014,
  author    = {Embrechts, P. and Kl{\"u}ppelberg, C. and Mikosch, T.},
  title     = {Quantitative Risk Management: Concepts, Techniques and Tools},
  publisher = {Princeton University Press},
  year      = {2014}
}

@article{Gerber1970,
  author  = {Gerber, H. U.},
  title   = {On Additive Risk Models and Brownian Motion},
  journal = {Insurance: Mathematics and Economics},
  volume  = {7},
  number  = {4},
  pages   = {289--303},
  year    = {1970}
}

@book{Glasserman2003,
  author    = {Glasserman, P.},
  title     = {Monte Carlo Methods in Financial Engineering},
  publisher = {Springer},
  year      = {2003}
}

@inproceedings{Goodfellow_et_al2014,
  author    = {Goodfellow, I. and Pouget-Abadie, J. and Mirza, M. and others},
  title     = {Generative adversarial networks},
  booktitle = {Advances in Neural Information Processing Systems (NeurIPS)},
  volume    = {27},
  pages     = {2672--2680},
  year      = {2014}
}

@book{Goodfellow2016,
  author    = {Goodfellow, I. and Bengio, Y. and Courville, A.},
  title     = {Deep Learning},
  publisher = {MIT Press},
  year      = {2016}
}

@inproceedings{Higgins2017,
  author    = {Higgins, I. and Matthey, L. and Pal, A. and Burgess, C. and Glorot, X. and Botvinick, M. and Mohamed, S. and Lerchner, A.},
  title     = {$\beta$-VAE: Learning Basic Visual Concepts with a Constrained Variational Framework},
  booktitle = {Proceedings of the International Conference on Learning Representations (ICLR)},
  year      = {2017}
}

@book{Klugman2012,
  author    = {Klugman, S. A. and Panjer, H. H. and Willmot, G. E.},
  title     = {Loss Models: From Data to Decisions},
  publisher = {John Wiley \& Sons},
  series    = {Wiley Series in Probability and Statistics},
  edition   = {4th},
  year      = {2012}
}

@misc{Kingma_and_Welling2014,
  author       = {Kingma, D. P. and Welling, M.},
  title        = {Auto-Encoding Variational Bayes},
  howpublished = {arXiv preprint arXiv:1312.6114},
  year         = {2014}
}

@book{Ross2014,
  author    = {Ross, S. M.},
  title     = {Introduction to Probability Models},
  publisher = {Academic Press},
  edition   = {11th},
  year      = {2014}
}

@book{Schmidli2008,
  author    = {Schmidli, H.},
  title     = {Stochastic Control in Insurance},
  publisher = {Springer},
  year      = {2008}
}

@article{Silver_et_al2016,
  author  = {Silver, D. and Huang, A. and Maddison, C. J. and others},
  title   = {Mastering the game of Go with deep neural networks and tree search},
  journal = {Nature},
  volume  = {529},
  number  = {7587},
  pages   = {484--489},
  year    = {2016}
}

@misc{Schulman_et_al2017,
  author       = {Schulman, J. and Wolski, F. and Dhariwal, P. and Radford, A. and Klimov, O.},
  title        = {Proximal policy optimization algorithms},
  howpublished = {arXiv preprint arXiv:1707.06347},
  year         = {2017}
}

@book{Sutton_and_Barto2018,
  author    = {Sutton, R. S. and Barto, A. G.},
  title     = {Reinforcement Learning: An Introduction},
  publisher = {MIT Press},
  year      = {2018}
}

\end{document}